\documentclass[12pt]{article} 
\usepackage{jheppub}

\usepackage{amsmath,amsfonts,amsbsy,amssymb,array,accents,dsfont}
\usepackage{enumerate}
\usepackage{graphicx} 
\usepackage{caption} 
\usepackage{subcaption} 
\usepackage{amssymb}
\usepackage{upgreek}
\usepackage{bm}
\usepackage{slashed}

\let\includefigures=\iftrue
\let\useblackboard==\iftrue
\newfam\black

\newcommand{\rcite}{\cite}

\def\kappab{{\pmb\kappa}}

\def\xx{{\bf  x}}
\def\yy{{\bf  y}}
\def\flabel{{\ell}}

\def\sl{\text{sl}}
\def\su{\text{su}}
\def \sutil{{\widetilde{\text{su}}}}

\def\sltwo{\ensuremath{SL(2,\bR)}}

\def\sutwo{{SU(2)}}
\def\sutwotil{{\widetilde{SU}(2)}}
\def\uone{U(1)}

\def\tight#1{\! #1 \!}  

\def\({\left(}
\def\){\right)}
\def\[{\left[}
\def\]{\right]}

\def\ie{{i.e.}}
\def\eg{{e.g.}}

\def\eff{{\rm eff}}

\def\tot{{\rm tot}}
\def\lpl{\ell_{\rm pl}}

\def\lstr{\ell_{\textit{s}}}

\def\gstr{g_{\textit s}^{\;}}
\def\gstrsq	{g_{\textit s}^{2}}

\def\nfive{{n_5}}

\def\none{{n_1}}

\def\sfA{{\mathsf A}}
\def\sfB{{\mathsf B}}
\def\sfC{{\mathsf C}}

\def\sfF{{\mathsf F}}
\def\sfH{{\mathsf H}}
\def\sfK{{\mathsf K}}

\def\sfX{{\mathsf X}}

\def\sfZ{{\mathsf Z}}

\def\sfv{{\mathsf v}}

\DeclareMathSymbol{\medhatsym}{\mathord}{largesymbols}{"62} 

\DeclareMathSymbol{\medtildesym}{\mathord}{largesymbols}{"65}

\def\ytil{{\tilde y}}

\def\Ry{R_y}

\def\half{\frac12}

\def\hf{\coeff12}

\def\One{{\hbox{1\kern-1mm l}}}

\def\barray{\begin{array}}
\def\earray{\end{array}}
\def\be{\begin{equation}}
\def\ee{\end{equation}}
\def\bea{\begin{eqnarray}}
\def\eea{\end{eqnarray}}
\def\bal{\begin{align}}
\def\eal{\end{align}}
\def\nn{\nonumber}

\newcommand{\bR}{{\mathbb R}}
\newcommand{\bS}{{\mathbb S}}
\newcommand{\bT}{{\mathbb T}}

\newcommand{\bZ}{{\mathbb Z}}

\definecolor{cardinal}{rgb}{0.6,0,0}
\definecolor{darkgreen}{rgb}{0,0.4,0}
\definecolor{green}{rgb}{0,0.4,0}
\definecolor{golden}{rgb}{0.92, 0.7, 0}
\definecolor{midnight}{rgb}{0, 0, 0.5}
\definecolor{darkblue}{rgb}{0, 0, 0.7}

\numberwithin{equation}{section}

\mathchardef\mhyphen="2D

\def\cG{\mathcal {G}} \def\cH{\mathcal {H}} \def\cI{\mathcal {I}}
\def\cJ{\mathcal {J}}  
\def\cM{\mathcal {M}} \def\cN{\mathcal {N}} 
  
 \def\cT{\mathcal {T}} 
 \def\cW{\mathcal {W}}

\def\one{{\hbox{\kern+.5mm 1\kern-.8mm l}}}
\def\zero{{\hbox{0\kern-1.5mm 0}}}

\newcommand{\ket}[1]{{\,| {#1} \rangle}}

\def\id{\textrm{id}}

\def\id{{1 \kern-.28em {\rm l}}}

\def\journal#1&#2(#3){\unskip, \sl #1\ \bf #2 \rm(19#3) }
\def\andjournal#1&#2(#3){\sl #1~\bf #2 \rm (19#3) }

\def\ie{{\it i.e.}}
\def\eg{{\it e.g.}}

\def\sst{\scriptscriptstyle}

\def\half{\frac12}
\def\hf{{\textstyle\half}}
\def\ket#1{|#1\rangle}

\def\One{{1\hskip -3pt {\rm l}}}

\catcode`\@=11
\def\slash#1{\mathord{\mathpalette\c@ncel{#1}}}
\def\underrel#1\over#2{\mathrel{\mathop{\kern\z@#1}\limits_{#2}}}

\catcode`\@=12

\def\ket#1{\left| #1\right\rangle}

\def\exp{{\rm exp}}

\def\ie{{\it i.e.}}
\def\eg{{\it e.g.}}

\def\mbar{{\bar m}}

\title{
{
\texorpdfstring{${\bf AdS}_{\bf 3}$}{}}'s with and without BTZ's
}

\author{Emil J. Martinec}

\affiliation{
\vskip 0.01cm
Kadanoff Center for Theoretical Physics and Enrico Fermi Institute\\ University of Chicago, Chicago IL 60637\\ 
}

\abstract{%
This companion paper to arXiv:2109.00065 explores the interpretation of $AdS_3/CFT_2$ duality for $k=(R_{AdS}/\lstr)^2<1$ in terms of ``non-critical little string theory''. 
We review the underlying fivebrane structure of 1/2-BPS backgrounds for string theory in both the critical dimension and for non-critical models.
D-branes bound to the fivebranes in these backgrounds are associated to nonabelian little strings, and flesh out a picture of the correspondence transition at $k=1$ wherein the nonabelian little string excitations lose their Hagedorn entropy, allowing the Hagedorn entropy of fundamental strings to dominate the asymptotic density of states for $k<1$ instead of BTZ black holes.
We then discuss the application of these models as approximations to and probes of black hole dynamics and thermodynamics, as well as associated information puzzles.  In particular, because the correspondence transition reveals the interior of an evaporating black hole, it serves as a sensitive probe of any proposal regarding the composition of that interior. 
}

\begin{document}
\hypersetup{pageanchor=false}
\begin{titlepage}
\maketitle
\thispagestyle{empty}
\end{titlepage}
\hypersetup{pageanchor=true}
\pagenumbering{arabic}

\thispagestyle{empty}

\vskip 1cm
\hrule


\section{Introduction} 
\label{sec:intro}

A new species of holographic duality was recently discovered in~\rcite{Balthazar:2021xeh}, which describes $AdS_3$ string theory in the regime $(R_{\it AdS}/\lstr)^2\equiv k<1$.
The stringy gravitational side of the duality involves $AdS_3$ times a squashed three-sphere $\bS^3_\flat$, while the corresponding CFT is a deformation of the symmetric product of $\bR_\rho\times\bS^3_\flat$.  The target space of this conformal sigma model describes the space {\it transverse} to the NS5-branes and fundamental (F1) strings that generate the gravitational background in a particular near-source limit.  In particular, the radial direction parametrized by $\rho$ is in full view.
This duality is thus quite different from the familiar example of the critical dimension duality between $AdS_3\times\bS^3\times\bT^4$ on the one hand, and a $CFT_2$ which is a deformation of the symmetric product of $\bT^4$ on the other hand.  In this more familiar example, the target space of the conformal sigma model describes the space {\it internal} to the background fivebranes; as such, determining the radial location of a particular CFT excitation in the dual geometry has proven quite difficult.

This distinguishing feature seems intimately related to the fact that the latter CFT describes the internal microstates of BTZ black holes in the critical dimension, while there are no black holes in the spectrum of the $k<1$ theory~\rcite{Giveon:2005mi}~-- the CFT instead describes a gas of fundamental strings in the fivebrane throat.

This companion article to~\rcite{Balthazar:2021xeh} elaborates on the interpretation of the $k<1$ models given there in terms of little string theory.  After reviewing the little string interpretation of the BTZ entropy in the critical dimension theory in section~\ref{sec:critdim}, in sections~\ref{sec:NS5} and~\ref{sec:Dp} we review evidence for this interpretation provided by the D-brane spectrum~\rcite{Martinec:2019wzw} in the regime somewhat below the threshold for black hole formation, where the stringy nature of the fivebrane source is well-understood using techniques of worldsheet string theory~\rcite{Martinec:2020gkv}.  

It is often said that AdS/CFT duality involves a geometric transition wherein branes dissolve into flux of the antisymmetric tensor fields they source, and thus disappear from the bulk description.  But in the case of $AdS_3/CFT_2$ duality where the background has NS flux, the background NS5-branes are solitons of closed string theory.  A detailed analysis of the worldsheet description of $\half$-BPS states reveals a stationary configuration of fivebranes in the background.  This configuration is determined by the state of the fundamental (F1) strings dissolved in the fivebranes.  This string condensate is hidden in worldsheet structures that are non-perturbative in the string scale $\alpha'=\lstr^2$ and thus invisible to supergravity.

By working near the BPS bound somewhat below the black hole threshold, with the NS5 sources slightly separated by their back-reaction to the F1 condensate they are carrying, we can keep strongly coupled nonabelian little string dynamics at bay, and have access to the near-source structure using the tools of perturbative string theory.  

D-branes in these backgrounds have all the hallmarks of nonabelian little strings made massive by separating the fivebranes slightly onto their Coulomb branch.  We review their structure in section~\ref{sec:Dp}, following~\rcite{Martinec:2019wzw,Martinec:2020gkv}.  In particular, these D-branes become light and are expected to control the infrared dynamics precisely when fivebranes collide and begin to restore nonabelian fivebrane dynamics.  Altogether, one arrives at a picture of the Hawking-Page transition in the NS5-F1 system as a deconfinement transition for nonabelian little strings, similar to all the other gauge/gravity dualities in which this transition involves a deconfinement of nonabelian gauge degrees of freedom.  The advantage here is that, because the NS5-branes are solitonic and we have some access to nonperturbative (in $\alpha'$) aspects of the worldsheet description, we can see all of this happening on the gravity side of the duality.

Section~\ref{sec:noncrit} then repeats this worldsheet analysis for non-critical models with $k<1$ as well as a series of models with $1\le k<2$, following~\rcite{Brennan:2020bju}.  We find that the D-brane avatars of the nonabelian little string have no transverse oscillations for $k<1$, but do in the models with $k>1$.  We take this as evidence that the Hagedorn entropy of little strings disappears for $k<1$, and that this vast reduction in the phase space available to the little string is responsible for the disappearance of the BTZ spectrum for $k<1$.  The D-branes in the $k<1$ models are also unstable to rapidly decaying into closed string radiation, thus bolstering the rationale for a dual CFT description in terms of the configuration space of fundamental strings rather than that of little strings.  

Altogether, one arrives at a heuristic explanation for why both the $k<1$ models and the critical dimension models have the form of a deformed symmetric product~-- both are describing a configuration space of string oscillations, however one is the fundamental string propagating in the fivebrane throat while the other is the little string bound to and propagating along the fivebrane.%
\footnote{The symmetric product in the critical dimension describes the corner of the moduli space where the effective description has a single fivebrane~\rcite{Seiberg:1999xz,Larsen:1999uk}.}

The value $k=1$ marks the correspondence transition, below which black holes are absent from the spectrum.  But one may ask to what extent the $k<1$ models approximate the spectrum and dynamics of black holes as one approaches the correspondence transition from below, due to the spacetime CFT having a deformed symmetric product structure very similar to a particular limit of the critical dimension theory.    Section~\ref{sec:BHapprox} is devoted to a discussion of this question.

In section~\ref{sec:EPRnotER} we extract some lessons about the nature of holography and the role of the correspondence transition in black hole evaporation and the information problem.  The absence of BTZ states for the $k<1$ models calls into question whether a high degree of quantum entanglement is always associated to the formation of wormholes, providing a counterexample to the notion that a holographic theory builds such geometries through its entanglement structure.  

We also consider circumstances in which a black hole passes through the correspondence transition during the course of evaporation, after which it becomes a gas of highly excited strings and branes~-- a system with no horizon.  Whatever remains in the black hole interior at the time of the transition is then revealed to the outside observer.  We consider a gedanken experiment in which the black hole passes through the correspondence transition around the Page time.  If the black hole interior teleports itself to some ``island'' in the state space of the Hawking radiation at the Page time, then dramatically different outcomes occur depending on whether the correspondence transition or the Page time comes first.  If on the other hand the black hole is some ``fuzzball'' (of highly excited strings and branes), gradually radiating like any other blackbody would, then nothing special happens at the Page time.  The correspondence transition thus discriminates between these two scenarios.


\section{Little string theory on \texorpdfstring{$\bS^1_x\times\bT^4$}{} }
\label{sec:critdim}

An intriguing early result on on black holes in string theory was the observation of~\rcite{Horowitz:1996ay} 
interpreting black hole thermodynamics in string theory on $\bS^1_x\times \bT^4$ in terms of a brane/antibrane gas, showing that their energy and entropy can be written as
\begin{align}
E_{\scriptscriptstyle\rm BH} &=  
\frac{R_x V_{\bT^4}}{\gstrsq}\big(n_5\!+\! n_{\bar 5}\,\big)
+ R_x \big({n_1}\!+\!{ n_{\bar 1}}\,\big)
+\frac1{R_x}\big({n_p}\!+\!{ n_{\bar p}}\,\big)
\nn\\[.2cm]
S_{\scriptscriptstyle\rm BH} &= 2\pi 
\big(\sqrt{n_5}\!+\!\sqrt{ n_{\bar 5}}\,\big)
\big(\sqrt{n_1}\!+\!\sqrt{ n_{\bar 1}}\,\big)
\big(\sqrt{n_p}\!+\!\sqrt{ n_{\bar p}}\,\big)
\end{align}
where $n_i^{~},n_{\bar \imath}^{~}$ are interpreted as effective numbers of NS5-branes, fundamental strings (F1), and momentum (P) quanta, as well as the corresponding antibranes and oppositely directed momentum quanta.  
Now of course black holes are not weakly interacting brane gases (though they do become such gases in stringy regimes~\rcite{Horowitz:1996nw}).  Nevertheless, the above expressions suggest the idea that black hole microstates are indeed complicated and highly entropic brane bound states.

Various decoupling limits suppress antibranes of a particular type by suppressing brane/antibrane pairs.  This suppression is accomplished by dialing the parameters that control the free energy cost of branes of that particular type.  For instance, the fivebrane decoupling limit, which is the focus of this paper, takes $\gstr\to0$ holding the net brane charge $n_5^{~}\tight-n_{\bar 5}^{~}$ and the energy above extremality fixed; then $n_{\bar 5}^{~}$ is forced to zero, leaving fundamental string winding/anti-winding and momentum/anti-momentum excitations intact.  It was pointed out in~\rcite{Maldacena:1996ya} that the resulting thermodynamics could be rewritten as that of an effective string gas (see also~\rcite{Larsen:1997ge,Martinec:1999gw})
\begin{align}
\label{blackNS5}
E^2 &= \Big(\frac p{R_x}+\frac{w R_x}{\alpha'_\eff}\Big)^2 + \frac 4{\alpha'_\eff}\, N_L
= \Big(\frac p{R_x}-\frac{w R_x}{\alpha'_\eff}\Big)^2 + \frac 4{\alpha'_\eff} \, N_R
\nn\\[.2cm]
S &= 2\pi \sqrt{N_L}+ 2\pi \sqrt{N_R}
\end{align}
with $\alpha'_\eff = \nfive\alpha'$, $p=n_p^{~}-n_{\bar p}^{~}$, and $w=n_5^{~}(n_1^{~}-n_{\bar 1}^{~})$.  The factor of $\nfive$ in the effective string's net winding number is understood from the idea that a fundamental string fractionates into $\nfive$ {\it little strings} when it is absorbed into a stack of $\nfive$ NS5-branes~\rcite{Maldacena:1996ya,Dijkgraaf:1997ku}.

The success of a free string expression for the entropy is not meant to indicate that the little string is in any sense weakly interacting; far from it.  Rather, this expression captures the average density of states, and indicates that the effect of interactions does not redistribute that density of states by an order one amount.

\begin{figure}[ht]
\centering
\includegraphics[width=.4\textwidth]{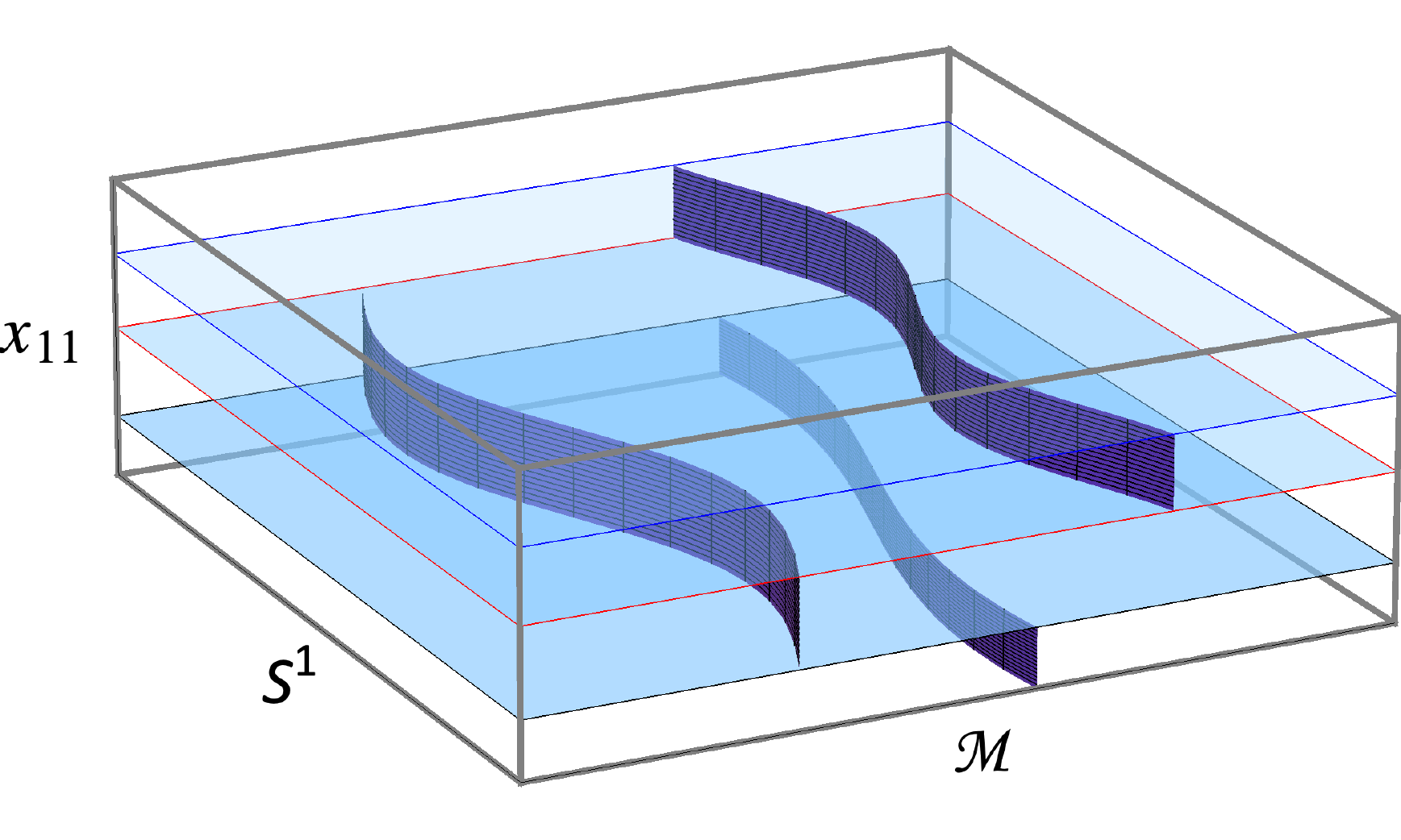}
\caption{\it The M-theory lift of a IIA NS5-F1 bound state, with the fivebranes wrapping $\bS^1\times\cM$ (with say $\cM=\bT^4$) and the onebranes wrapping $\bS^1$.}
\label{fig:wiggles}
\end{figure}

There are various heuristic pictures of this fractionation phenomenon.  One that we will find useful is to lift type IIA NS5-branes to M5-branes in M-theory, and separate the fivebranes as well in the additional circle of radius $R_{11}$ that arises in this limit (see figure~\ref{fig:wiggles}).  The fundamental string lifts to an M2-brane wrapped around the circle.  

%
%

When an M2-brane encounters the stack of M5-branes that are coincident in the original ten dimensions but separated along a transverse M-theory circle, it can break into $n_5$ strips, whose average tension when viewed as an effective string is $\nfive$ times smaller than the fundamental string tension.  The fractionated strips can then separate along the fivebrane worldvolume but are pinned in the transverse space to the location of the fivebranes.  These stringlike objects descend to the little strings in the type IIA string theory limit, when the radius $R_{11}$ of the M-theory circle is shrunk to sub-Planckian size.%
\footnote{In the IIA limit one expects quantum effects to smear out the location of the fivebranes along the M-theory circle since it has shrunk below any physical resolution (\ie\ Planck) scale; but the idea is that there is a sense in which the fivebranes are on average separated on the circle, and so the picture is perhaps valid in some average sense.}

The resulting little string theory compactified on $\bS^1_x\times \bT^4$ is dual to the linear dilaton geometry of the throat of coincident fivebranes.  One can take a further decoupling limit $R_x\to\infty$ in the presence of F1 winding charge on $\bS^1_x$, in which the asymptotic spacetime geometry becomes $AdS_3$, and the IR limit of the little string theory is a ${\it CFT}_2$ on $\bR_t\times \bS^1_x$.  Taking this limit in the thermodynamics, the energetic cost of ${\rm F1}$-$\rm\overline{F1}$ pairs forces $n_{\bar 1}^{~}\to 0$.  We write 
\be
E = \frac{\varepsilon}{R_x}+\frac{\none R_x}{\alpha'}
\ee 
in order to measure energy relative to the rest energy of the winding strings, which are now part of the background.  We then find (dropping terms that vanish as $R_x\to\infty$)
\begin{align}
\label{BTZspec}
\varepsilon &= p + \frac{2N_L}{\nfive\none} = -p + \frac{2N_R}{\nfive\none}
\nn\\[.2cm]
S &= 2\pi \sqrt{N_L}+ 2\pi \sqrt{N_R} = 2\pi \sqrt{\nfive\none(\varepsilon+p)/2} + 2\pi \sqrt{\nfive\none(\varepsilon-p)/2} ~,
\end{align}
which is indeed the asymptotic density of states of a 2d CFT with central charge $6\nfive\none$ at left and right conformal dimensions $(\varepsilon\pm p)/2$.  We conclude that the limiting BTZ entropy is accounted for by little strings winding $\bS^1_x$ and oscillating along $\bT^4$, with central charge $c=6$ per unit little string winding (\ie\ with a total winding $\nfive\none$, and thus total central charge $6\nfive\none$).

These sorts of global considerations give us a hint as to the degrees of freedom underlying black hole entropy, but they don't give any clues about the internal structure of the bound state and in particular the nature of the horizon in the strong coupling regime where the gravity approximation applies.  We can, however, explore  the regime near but slightly below the threshold of black hole formation, and ask whether we can see these brane degrees of freedom get activated as we near the black hole threshold.  The idea is that as we approach the threshold from below, there are deep redshifts but as yet no horizon; the nonabelian little string degrees of freedom that are activated in the black hole phase should be energetically suppressed below threshold, but becoming lighter and lighter the closer we are to the threshold.  Fortunately, in the NS5-F1 system, there is a rich structure of $\half$-BPS states below threshold, which are under good control and have limits where they approach the black hole threshold.  
The frame with only NS fluxes is somewhat unique in that one can analyze them effectively using the tools of worldsheet string theory and thereby gain an understanding of effects beyond supergravity in the $\alpha'$ expansion.
We turn next to a review of these states.


\section{NS5-brane backgrounds near the black hole threshold}
\label{sec:NS5}

In addition to the effective picture of thermal states outlined above, we also have some understanding of vacuum structure, as well as a rich collection of $\half$-BPS states.  We begin with the vacuum of coincident fivebranes.

The geometry of $\nfive$ coincident NS5-branes is given by%
\footnote{All metrics will be written in string frame.}
\begin{align}
\label{nfivegeom}
ds^2 &= \Big( -dt^2 + d\ytil^2 + ds_\cM^2 \Big)_{||} + \sfH\Big[ dr^2 + r^2\Big( d\theta^2+ \sin^2\!\theta\,d\phi^2 + \cos^2\!\theta\, d\psi^2 \Big)\Big]_{\perp}
\\[.2cm]
H_{\theta\phi\psi} &= \epsilon_{\theta\phi\psi}^{~~~r}\,\partial_r\sfH
~~,~~~~
e^{2\Phi} = \sfH
~~,~~~~
\sfH = \gstrsq + \frac{\nfive\,\alpha'}{r^2}  ~.
\end{align}
In the fivebrane decoupling $\gstr\to 0$, the location of the opening to the asymptotically flat region at the top of the throat is sent off to $r\to\infty$.  Note that although the asymptotic $\gstr$ is being sent to zero, the string coupling throughout the throat stays finite in the limit, so that there is a nontrivial interacting string theory, and in particular the coupling still diverges as $r\to0$.  In this fivebrane decoupling limit, $\rho=\log r$ is a free field with a linear dilaton, and the angular sphere is described by an $\sutwo$ WZW model at level $\nfive$.  Thus the worldsheet theory is exactly solvable~\rcite{Callan:1991at}, but the theory is singular from the point of view of string perturbation theory~-- strings can freely propagate to the strong coupling region $r\to 0$ where one loses perturbative control.  In order to use worldsheet methods, one needs an infrared regulator that keeps strings away from strong coupling.

\subsection{IR regulator \#1: The Coulomb branch of NS5-branes}
\label{sec:NS5Coul}

\begin{figure}[ht]
\centering
  \begin{subfigure}[b]{0.33\textwidth}
    \includegraphics[width=\textwidth]{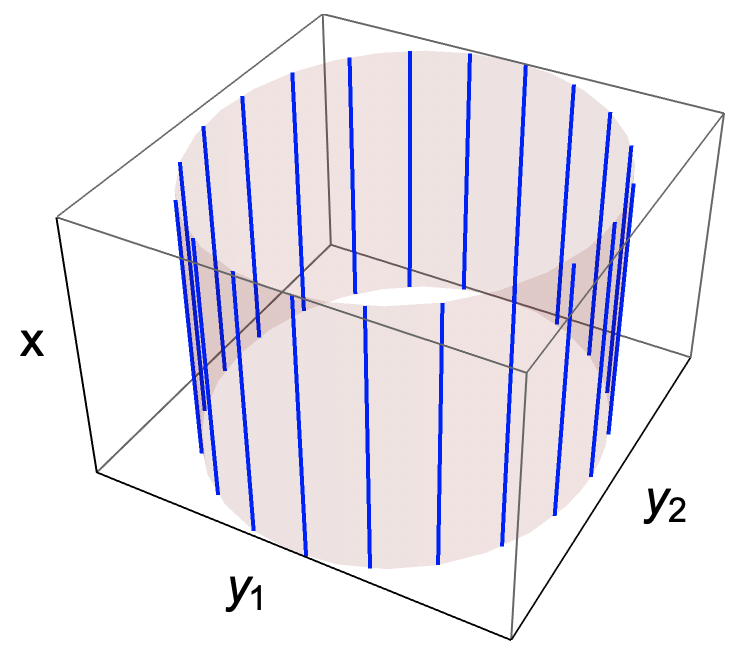}
    \caption{ }
    \label{fig:Coulsource}
  \end{subfigure}
\hskip 1.5cm
  \begin{subfigure}[b]{0.33\textwidth}
    \includegraphics[width=\textwidth]{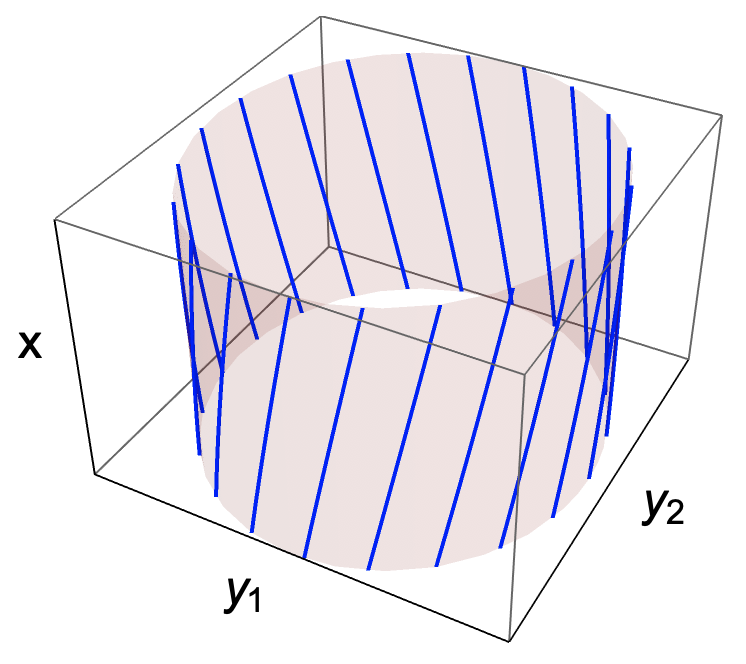}
    \caption{ }
    \label{fig:STsource}
  \end{subfigure}
\caption{\it 
(a) Source configuration for NS5-branes on the Coulomb branch.\\
(b) Source configuration for a circular NS5-P ``supertube'', T-dual to an NS5-F1 source.
}
\label{fig:sources}
\end{figure}

There are two ways to address this issue, which we consider in turn.  The first IR regulator we will discuss is to separate the fivebranes slightly in their transverse space $\bR^4$, so that the harmonic function $\sfH$ becomes
\be
\label{n5coul}
\sfH = \sum_{\ell=1}^\nfive \frac{\alpha'}{|\yy-\yy_\ell |^2}  ~.
\ee
Naively, this doesn't improve the situation, simply replacing one coincident brane singularity by several individual ones.  However, the level $\kappa_\tot=\nfive$ of the supersymmetric $\sutwo$ WZW model describing the angular space in~\eqref{nfivegeom} gets a contribution $\kappa=\nfive\tight-2$ from the bosonic $\sutwo$ WZW model and $\kappa\tight=2$ from its fermionic superpartners.  Thus there is no unitary worldsheet description of local string dynamics in the throat of a single NS5-brane.%
\footnote{Recently, a worldsheet theory has been proposed to describe a single fivebrane in the presence of a macroscopic number of wound fundamental strings~\rcite{Eberhardt:2018ouy}.  We are not aware of an extension of this description where there is a fivebrane but no wound strings; and furthermore, in this description the wound strings have a single allowed radial wavefunction, whose radial profile is essentially constant, so there is still no local string dynamics in the single fivebrane throat.}
The way the worldsheet theory accommodates this fact is to not allow fundamental strings to penetrate into the throat of an isolated fivebrane at low energy~\rcite{Giveon:1999px,Giveon:1999tq}.   This property has been studied in detail for the particular configuration where the fivebranes are separated in a circular array in say the $y^1$-$y^2$ plane (see figure~\ref{fig:Coulsource}) 
\be
\label{circarray}
y_\ell^1+i y_\ell^2 = a\,e^{2\pi i\ell/\nfive} ~.
\ee
This configuration is special in that it still admits an exactly solvable worldsheet description~\rcite{Giveon:1999px,Giveon:1999tq}.

The version of this exact worldsheet theory that will be useful for us is a presentation in terms of a gauged WZW model~\rcite{Israel:2004ir,Martinec:2017ztd}
\be
\label{ngwzwCoul}
\frac\cG\cH = \frac{\sltwo\times\sutwo}{\uone_L\times \uone_R}
\ee
where $\uone_L\times\uone_R$ is generated by the left- and right-moving null currents
\be
\label{nullcurrCoul}
\cJ = J^3_\sl + J^3_\su
~~,~~~~
\bar\cJ = \bar J^3_\sl - \bar J^3_\su ~.
\ee
Gauging these currents and integrating out the gauge field leads to a geometry which when written in terms of the natural Euler angle coordinates on the $\sltwo\times\sutwo$ group manifold%
\footnote{The metric and NS $B$-field in Euler angles on the angular three-sphere $\sutwo$ is given in~\eqref{nfivegeom}.  For $\sltwo$, substitute $\theta\to i\rho$, $\phi\to\sigma$, $\psi\to\tau$; it is convenient to fix $\tau\tight=\sigma\tight=0$ as a gauge choice.  See~\rcite{Martinec:2017ztd,Martinec:2018nco,Martinec:2019wzw,Martinec:2020gkv} for details.
}
takes the form
\begin{align}
\label{NS5 Coul}
ds^2 &= \Bigl(-dt^2 +dx^2 +ds_{\scriptscriptstyle\mathbf T^4}^2 \Bigr)
+ n_5\Bigl[d\rho^2 + d\theta^2 + \frac{1}{\Sigma_0}\Bigl(
{\cosh}^2\!\rho\sin^2\!\theta \,d\phi^2 + {\sinh}^2\!\rho\cos^2\!\theta \,d\psi^2\Bigr)\Bigr],
\nn\\[8pt]
B   &= \frac{n_5 \cos^2\theta \cosh^2\rho}{\Sigma_0} \, d\phi \wedge d\psi
\,,\qquad
e^{-2\Phi} = \frac{a^2\Sigma_0}{\nfive}  \,, \qquad 
\Sigma_0 \;\equiv\; {\sinh^2\!\rho + \cos^2\!\theta} \,.
\end{align}
This geometry is just what one would get from smearing the Coulomb branch NS5 sources over the circle along which they are uniformly arrayed~\rcite{Sfetsos:1998xd}, with the coordinate map
\be
\label{bipolar}
y^1+iy^2 = a \, \cosh\rho\, \sin\theta\, e^{i\phi}
~~,~~~~
y^3+iy^4 = a \, \sinh\rho\, \cos\theta\, e^{i\psi} ~.
\ee
From the point of view of the effective supergravity theory, the geometry is still singular along the circle $\rho=0,\theta=\pi/2$ parametrized by $\phi$, as one sees from the the vanishing of $\Sigma_0$ in~\eqref{NS5 Coul};
however, the worldsheet CFT is completely nonsingular~-- its correlation functions are simply a projection of the nonsingular correlation functions of the parent WZW model onto the subset that are neutral under the currents~\eqref{nullcurrCoul}.  Thus the apparent singularity is a fake, at least from the point of view of low-energy strings~-- it is effectively resolved by $\alpha'$ corrections.  The effective coupling in the theory is set by the value of the dilaton away from this source ring but in the cap of the geometry, say at the center of the source ring at $\rho=\theta=0$.  Note that the effective supergravity geometry~\eqref{NS5 Coul} does not resolve the individual fivebrane locations; rather the source appears to be smeared all along the circle.  Nevertheless, the exact worldsheet theory keeps track of the precise location of the fivebrane sources.

A physical picture of what is happening is provided by the FZZ dual description of the near-source region~\rcite{Giveon:1999px,Giveon:1999tq,Martinec:2020gkv}.  This duality initially arose in the present context as a quantum equivalence between the supersymmetric $\frac\sltwo\uone$ gauged WZW model, and $N=2$ supersymmetric Liouville field theory~\rcite{Giveon:1999px}.  Ultimately, though, it derives from a certain quantum equivalence of $\sltwo$ current algebra representation theory, as explained in~\rcite{Maldacena:2000hw,Chang:2014jta,Giveon:2015cma,Martinec:2020gkv}.

In this quantum equivalence of 2d CFT's, the IR cap in the geometry~\eqref{NS5 Coul} is seen to be equivalent to a Liouville-like wall, described by the $N=2$ supersymmetric worldsheet superpotential
\be
\label{spotl}
\cW = \sfZ^\nfive - \lambda_0\, e^{-\nfive \sfX} = \prod_{\ell=1}^\nfive\Big(\sfZ - \mu_\ell\,e^{-\sfX}\Big)
~~,~~~~
\mu_\ell = \lambda_0^{1/\nfive}\,e^{2\pi i\ell/\nfive}  ~.
\ee
The $N=2$ Liouville field $\sfX$ has a linear dilaton for its real part $\rho=\Re(\sfX)$, which represents the radial direction away from the fivebrane source; the exponential wall keeps strings from pushing into the strong coupling region.  The circular array of fivebranes is coded in the zeros of this superpotential; the coupling $\lambda_0$ thus sets a scale related to the parameter $a$ in the geometrical description.  This worldsheet dual description captures information that is non-perturbative in $\alpha'$ relative to the sigma model description~\eqref{NS5 Coul}, such as the unsmeared locations of the fivebrane sources.

Thus the throats of isolated fivebranes cannot be probed by fundamental strings at low energy, because a wall prevents strings from reaching the region of strong coupling near such a fivebrane.  However, pushing the branes back together by sending $a\to0$ and hence $\lambda_0\to0$ makes the effective wall recede and strings are no longer kept away from the region of strong coupling near the fivebranes.  

Note also that the parameters $\mu_\ell$ are marginal couplings in the superpotential~\eqref{spotl}, related to the positions of the individual fivebranes $\yy_\ell$ in the $y^1$-$y^2$ plane.  One can push any pair $\ell,\ell'$ of fivebranes together by sending $\mu_\ell\to\mu_{\ell'}$, in which case the superpotential develops a flat direction and once again, along that flat direction the dilaton grows and perturbative strings can explore it, because the throat of two fivebranes has a unitary worldsheet description as outlined above.

Separating the fivebranes onto their Coulomb branch moduli space thus regulates the strong coupling IR dynamics of the nonabelian little string theory that lives inside coincident NS5-branes.


\subsection{IR regulator \#2: Adding fundamental strings}
\label{sec:NS5F1}

\begin{figure}[ht]
\centering
\includegraphics[width=.6\textwidth]{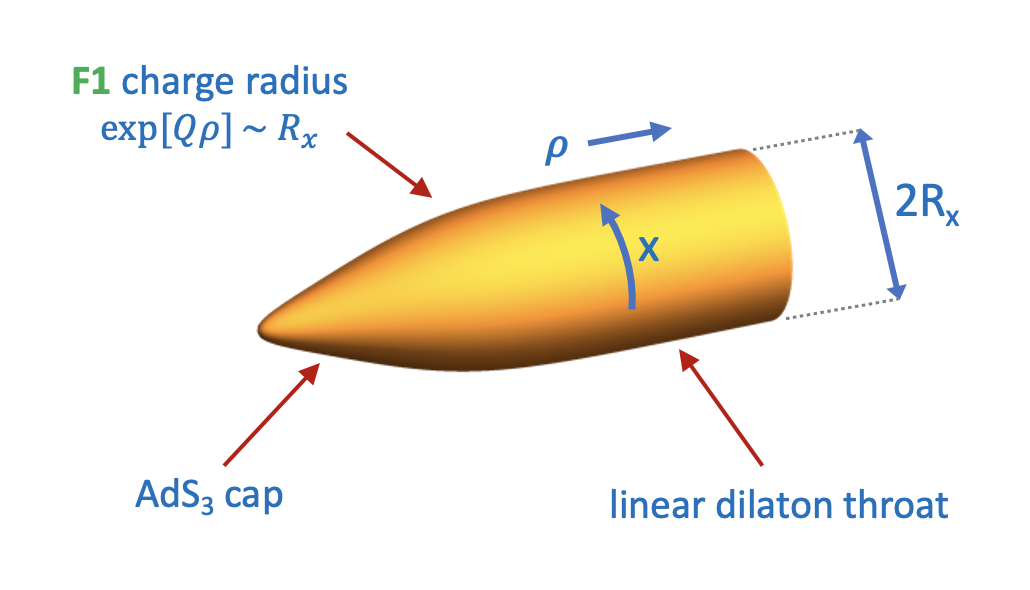}
\caption{\it Interpolating spatial geometry between $AdS_3$ in the IR and cylindrical $\bR_\rho\times\bS^1_x$ with a linear dilaton in the UV.
}
\label{fig:AdS2lindil}
\end{figure}

A second infrared regulator results from dropping into the throat $\none$ fundamental strings which wrap $\bS^1_x$.  Inside the fivebrane charge radius $r_{\rm F1}^{~}$, the linear dilaton throat of the NS5-branes rolls over into an infrared $AdS_3$ geometry, and the dilaton saturates to a constant $e^{2\Phi}\approx V_{\bT^4}^{~} n_5/n_1$; see figure~\ref{fig:AdS2lindil}.  The theory is weakly coupled in the IR if the F1 winding charge $\none$ is sufficiently large.

%
%

A worldsheet description of this transition between linear dilaton asymptotics and $AdS_3$ in the IR can be achieved via a suitable modification of the gauged WZW formalism described above~\rcite{Martinec:2017ztd}.  One simply generalizes the embedding of the gauge group to the $\uone_L\times \uone_R$ generated by
\be
\label{nullcurrNS5F1}
\cJ = J^3_\sl + J^3_\su - R_x (\partial t + \partial x )
~~,~~~~
\bar\cJ = \bar J^3_\sl - \bar J^3_\su  - R_x (\bar\partial t - \bar\partial x )  ~.
\ee
Gauging these currents in the WZW model leads to a somewhat more complicated geometry
\begin{align}
\label{smearedNS5F1metric}
ds^2 &= \Bigl( -du\, dv + ds_{\scriptscriptstyle\mathbf T^4}^2  \Bigr)
+ {n_5}\Bigl[ d\rho^2+ d\theta^2+\frac1\Sigma \Bigl( {\cosh}^2\!\rho\sin^2\!\theta \,d\phi^2 + {\sinh}^2\!\rho\cos^2\!\theta \,d\psi^2 \Bigr)\Bigr] 
\nonumber\\[.1cm]
& \hskip .6cm 
+ \frac{2R_x}{\Sigma} \Bigl(  {\sin^2\!\theta \, dt\,d\phi + \cos^2\!\theta \, dx\,d\psi}  \Bigr)
+\frac{R_x^2}{\nfive\Sigma} \Bigl[ \nfive \sin^2\!\theta \, d\phi^2 +  \nfive\cos^2\!\theta \, d\psi^2  
+ du\:\! dv \Bigr],
\nonumber\\[8pt]
B  &= \frac{ \cos^2\!\theta (R_x^2+\nfive\cosh^2\!\rho)}{\Sigma}  { d\phi\wedge d\psi - \frac{R_x^2}{n_5\Sigma} \, dt\wedge dx } \nonumber\\
& \hskip .6cm 
{}+\frac{R_x  \cos^2\!\theta}{\Sigma}  dt\wedge d\psi
+\frac{R_x  \sin^2\!\theta}{\Sigma}  dx\wedge d\phi~,
\nn\\[8pt]
e^{-2\Phi} & = \frac{n_1\Sigma}{R_x^2\,V_4} ~,\qquad~~~  
\Sigma = \frac{R_x^2}{\nfive} + \Sigma_0 \ ,
\end{align} 
where $u,v=t\pm x$.

This background has the advertised feature, that it has a linear dilaton structure for $e^{2\rho}\gg R_x^2/\nfive$, and an $AdS_3$ structure for $e^{2\rho}\ll R_x^2/\nfive$.  One way to see these asymptotics is to note that the $\sltwo$ current $J^3_\sl$ in the null current $\cJ$ behaves as $J^3_\sl\sim \hf e^{2\rho}\partial(\tau+\sigma)$.  Therefore at large $\rho$ the gauge orbits lie mostly along $\sltwo$, the gauging is approximately that of the Coulomb branch fivebranes, and $t,x$ parametrize an asymptotic cylinder; on the other hand, for small $\rho$, the gauge orbit is mostly along $\bR_t\times\bS^1_x$, which are gauged away leaving $AdS_3$ largely intact.  Note also that the $B$ field now contains, in addition to the magnetic flux represented by the first term in the first line, an electric flux represented by the second term in the first line.  This arises because the $H_3$ flux in the $\sltwo$ factor is no longer unphysical after tilting the orientation of the gauge orbits to partly lie along $\bR_t\times \bS^1_x$. 

The $AdS_3$ decoupling limit of this background takes $R_x\to\infty$.  Because spacetime fermions have periodic boundary conditions around $\bS^1_x$, the resulting background describes a state in the Ramond sector of the dual spacetime CFT; in fact one obtains the Ramond sector state of maximum R-charge, which is the spectral flow of the $\sltwo$ invariant state in the NS sector~\rcite{Maldacena:2000dr,Lunin:2001fv}.  This property explains why the geometry looks like (a large diffeomorphism of) global $AdS_3$ in the IR.%
\footnote{A suitable generalization of the gauge group embedding allows the realization of conical defect geometries $AdS_3\times \bS^3)/\bZ_q$, with R-charge a faction $1/q$ of the maximal value~\rcite{Martinec:2017ztd}.}

In fact, this second IR regulator has a kinship with the Coulomb branch regulator above~\rcite{Martinec:2017ztd}.  T-duality along $x$ converts the F1 charge to momentum charge P, and the fivebrane source configuration in fact still has the $\nfive$ fivebranes uniformly spread around a circle, whose radius is now dynamically determined by the tension of the fivebranes balanced against the angular momentum they carry; see figure~\ref{fig:STsource} (this balancing determines the value of the dilaton in the NS5-F1 frame).  Thus, the source fivebranes are indeed separated along this T-dual circle $\widetilde \bS^1_ x$.  This uniformly spiraling source generates a particularly symmetric geometry, which is amenable to exact solution on the worldsheet.

\begin{figure}[ht]
\centering
\includegraphics[width=.5\textwidth]{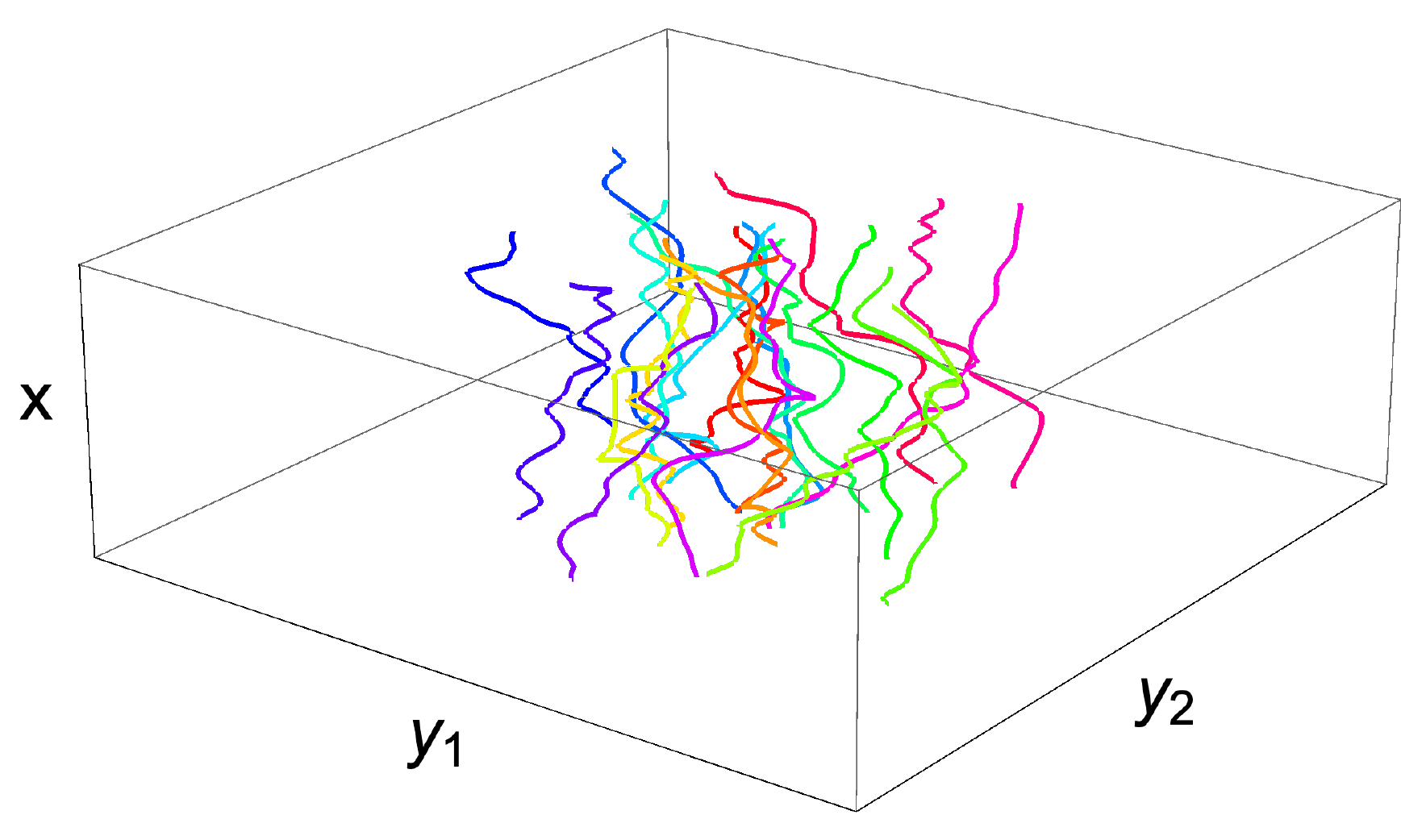}
\caption{\it A typical Lunin-Mathur source profile, with the joining of the fivebrane strands indicated by the continuously varying hue.  }
\label{fig:LMwiggles}
\end{figure}

The maximal R-charge Ramond ground state is of course one of a large ensemble of $\half$-BPS states.  The geometry sourced by these states can be given in terms of Green's function integrals for the coefficients in the supergravity fields; for intance, if only NS-NS fields are sourced, one has (see~\rcite{Mathur:2005zp,Skenderis:2008qn} for reviews)
\begin{align}
\begin{aligned}
\label{LMgeom}
ds^2 &\;=\; \sfK  ^{-1}\bigl[ -(d\tau+\sfA)^2 + (d\sigma + \sfB)^2 \bigr] + \sfH \, d\yy\!\cdot\! d\yy + ds^2_\cM \,,
\\[.2cm]
B &\;=\; 
\sfK  ^{-1}\bigl(d\tau + \sfA\bigr)\wedge\bigl(d\sigma+\sfB\bigr)  + \sfC_{ij}\, dy^i\wedge dy^j \,,
\\[.2cm]
e^{2\Phi} &\;=\; 
\frac{\sfH}{\sfK}  \,, \qquad~~~ d\sfC =  *_{\sst\perp} d\sfH \;, 
\qquad~~ d\sfB = *_{\sst\perp} d\sfA \,,
\end{aligned}
\end{align}
where again $\yy$ are Cartesian coordinates on the transverse space to the fivebranes.
The harmonic forms and functions appearing in this solution can be written in terms of a Green's function representation,
which in the $AdS_3$ decoupling limit takes the form
\begin{align}
\begin{aligned}
\label{greensfn}
\sfH  \,=\, \frac{1}{2\pi}\sum_{\ell=1}^\nfive\int\limits_0^{2\pi} \frac{d\tilde{v}}{|\xx-\sfF_\flabel(\tilde{v})|^2} ~, & \qquad~
\sfK   \,=\,  
1 + \frac{1}{2\pi}\sum_{\ell=1}^\nfive\int\limits_0^{2\pi} \frac{d\tilde{v} \, \dot\sfF_\flabel \tight\cdot \dot\sfF_\flabel}{|\xx-\sfF_\flabel(\tilde{v})|^2} \;,\\[.2cm]
\sfA \,=\, \sfA_i dy^i \,, \qquad
\sfA_i \,=\, & \frac{1}{2\pi}\sum_{\ell=1}^\nfive\int\limits_0^{2\pi} \frac{d\tilde{v} \,\dot \sfF^\flabel_i(\tilde{v})}{|\xx-\sfF_\flabel(\tilde{v})|^2}\;,
\end{aligned}
\end{align}
involving source profile functions $\sfF_\flabel(\tilde{v})$, $\ell=1,2,\dots,\nfive$, that describe the locations of the fivebranes in their transverse space (overdots denote derivatives with respect to $\tilde{v}$).  It is convenient to choose twisted boundary conditions $\sfF_\flabel(\tilde{v}+2\pi)=\sfF_{\sst \ell+1}(\tilde{v}) $, so that all the fivebranes are bound together into one long fivebrane, see figure~\ref{fig:LMwiggles}.  
For example, the choice
\be
\sfF_{\sst \ell}(\tilde{v}) = a\, \exp\Big[ \frac{i\tilde v}{\nfive \Ry} + \frac{2\pi i \ell}{\nfive} \Big]
\ee
is the source profile that generates the round ``supertube'' geometry of equation~\eqref{smearedNS5F1metric} and figure~\ref{fig:STsource}.
%
%

One now has a fancier version of the Coulomb branch story.  For profiles which only excite wiggles of the fivebranes in the $y^1$-$y^2$ plane, there is a proposed FZZ dual description~\rcite{Martinec:2020gkv} in terms of an $N=2$ superpotential 
\be
\label{ST spotl}
\cW = \prod_{\ell=1}^\nfive \Bigl( \sfZ \,e^{i\sfv/\nfive} - \mu_\ell (\sfv)\, e^{-\sfX} \Bigr) 
\ee
where the $\mu_\ell$ are simply related to the source profiles $\sfF_\ell(\tilde v)$.  When the $\mu_\ell$ are the $n_5^{\rm th}$ roots of unity (a common overall scale can be absorbed in a shift of $\sfX$), one has a spiral structure of zeros due to the adiabatically rotating phase multiplying $\sfZ$; allowing more general $\mu_\ell(\sfv)$ with the twisted boundary condition $\mu_\ell(\sfv+2\pi)=\mu_{\ell+1}(\sfv)$ reproduces the wiggly profile of the source in the geometrical construction of Lunin-Mathur, and exhibits the unsmeared source in the worldsheet theory.

\begin{figure}[ht]
\centering
\includegraphics[width=.8\textwidth]{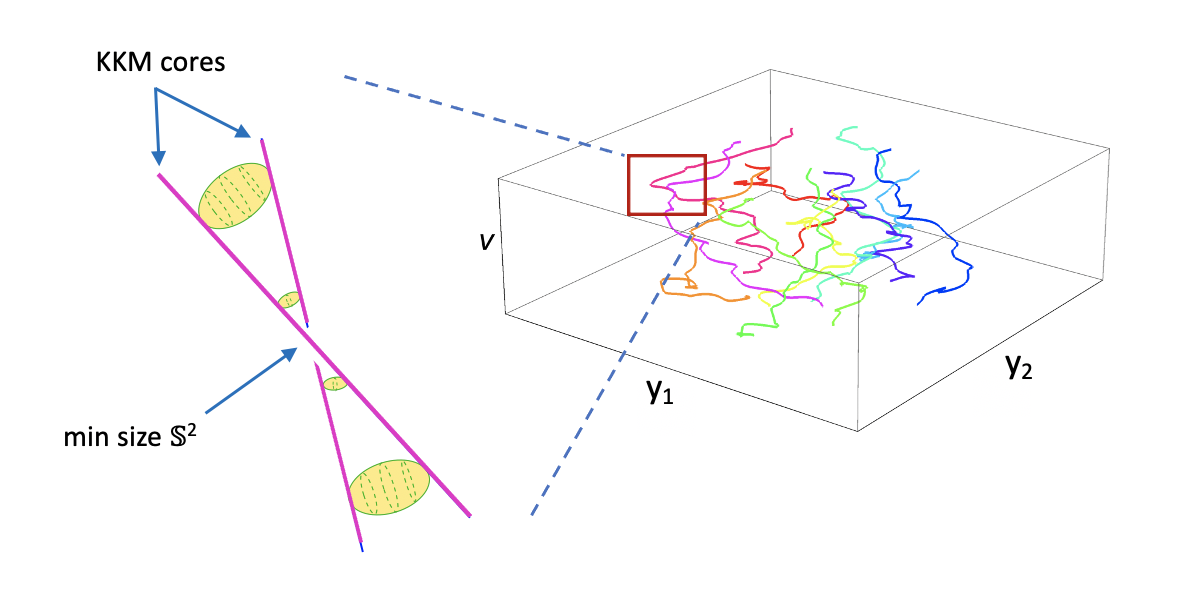}
\caption{\it Vanishing cycles appear when the Lunin-Mathur source profile self-intersects.  
}
\label{fig:singularity}
\end{figure}

Singularities arise when the profile self-intersects.  In the geometry~\eqref{LMgeom}, the source profile $\sfF(\tilde v)$ specifies the location of a KK monopole core transverse to the source contour.%
\footnote{The tilted source profile of figure~\ref{fig:STsource} has a component transverse to the circle $\bS^1_x$ that is being T-dualized to go from NS5-P to NS5-F1, and this component of NS5 charge then T-dualizes into KKM charge.  This is a dipole charge because its orientation rotates as one travels around the $\phi$ circle at $\rho=0,\theta=\frac\pi2$, such that the net KKM charge vanishes.}    
When the profile approaches a self-intersection, there is a minimal size holomogical $\bS^2$ in the geometry whose polar direction is the interval between the two points of closest approach of the contour, and whose azimuthal direction is the fibered circle of the KK monopole.  The size of this cycle vanishes in the limit where the profile actually self-intersects, see figure~\ref{fig:singularity}.
%
%

In the FZZ dual description, this self-intersection corresponds to a point along $\sfv$ where two zeros of the superpotential~\eqref{ST spotl} collide, for instance $\mu_\ell(\sfv)$ and $\mu_{\ell'}(\sfv)$; again the superpotential degenerates as it did on the Coulomb branch, a flat direction opens up, along which strings are no longer walled off from strong coupling in this direction in field space.

All of this structure reinforces the picture of a brane gas underlying the density of states.  Below the black hole threshold, we can enumerate the $\half$-BPS spectrum as the phase space of a wiggly fivebrane source carrying KK-dipole charge.  
When we put energy into the system, it starts to explore this phase space, until it finds a self-intersection of the wiggly fivebrane.  The formation of little strings will trap the intersecting fivebranes and start the black hole formation process.
A deep throat opens up in this wiggly geometry at the intersection.
In the black hole phase that is emerging, the phase space is dominated by that of nonabelian little string excitations.  The manifestation of these nonabelian little strings in perturbative string theory below the transition, and their emergence as light excitations as we approach the black hole threshold, is the subject we turn to next.


\section{Massive little strings as D-branes}
\label{sec:Dp}

The heuristic picture of the little string depicts it as a fractionated fundamental string dissolved in the fivebrane worldvolume, as discussed in section~\ref{sec:critdim} using the M-theory description.  

\begin{figure}[ht]
\centering
\includegraphics[width=.9\textwidth]{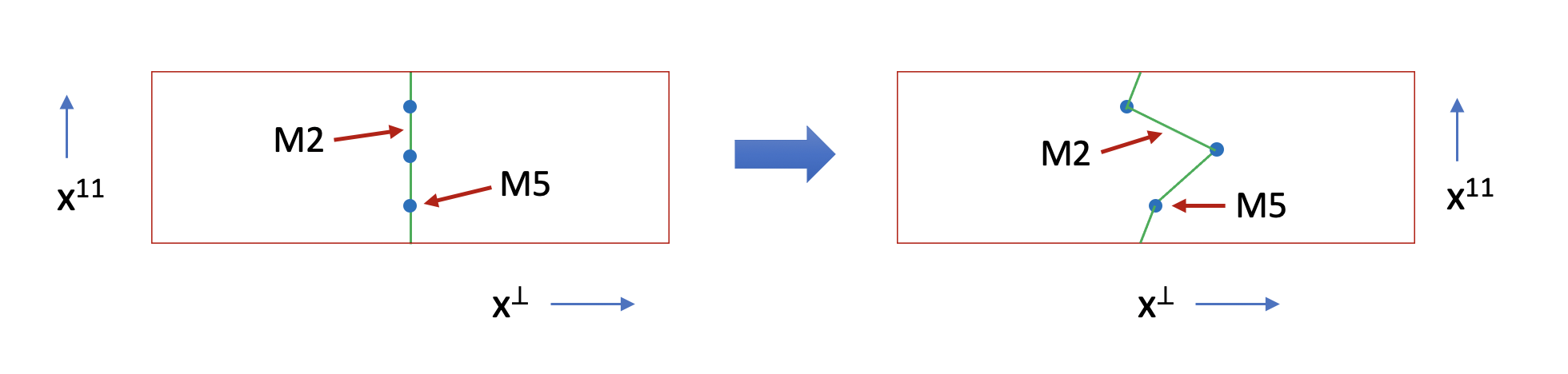}
\caption{\it Cartoon of fundamental string fractionation along fivebranes as seen from the M-theory picture, when the fivebranes are separated only along the M-theory circle; and the corresponding process when the fivebranes are also separated in another transverse dimension.  The M2-strips are generically separated in a direction along the M5-branes transverse to the plane of the figure, as depicted in figure~\ref{fig:wiggles}, and so don't necessarily reconnect immediately and lift off the M5-branes.
}
\label{fig:fractionation}
\end{figure}

Now consider separating the M5-branes in their transverse $\bR^4$ in ten dimensions as well.  Their reduction to type IIA will now realize the M2-brane strips as D2-branes stretching between the fivebranes, provided the separation in these directions is much larger than the Planck scale; see figure~\ref{fig:fractionation}.  One might then regard such D-branes as the avatars of little strings, made massive by moving the fivebranes onto their Coulomb branch.  It is these nonabelian ``W-strings'' that become light and restore the nonabelian little string theory in the limit when the fivebranes are pushed back together in their transverse space.

%
%

This picture survives the fivebrane decoupling limit, where we can study D-branes in the fivebrane throat~\rcite{Israel:2005fn,Elitzur:2000pq,Eguchi:2000cj,Martinec:2019wzw}.  In the gauged WZW model, the relevant D-branes are characterized by conjugacy classes in the numerator group (\eg\ two-spheres in $\sutwo$, and hyperboloids in $\sltwo$; see~\rcite{Martinec:2019wzw} and references therein).  The azimuthal direction of this $\bS^2$ is the Euler angle $\psi$, while the polar direction is a line across the disk parametrized by $\theta,\phi$.  The size of the $\bS^2$ branes in $\sutwo=\bS^3$ is dictated by flux quantization.  This D-brane in the $\sutwo$ part of the numerator group is tensored with an $\sltwo$ D-brane associated to the conjugacy class of the identity, smeared along the timelike direction $\tau$ in $\sltwo$.

\begin{figure}[ht]
\centering
\includegraphics[width=.5\textwidth]{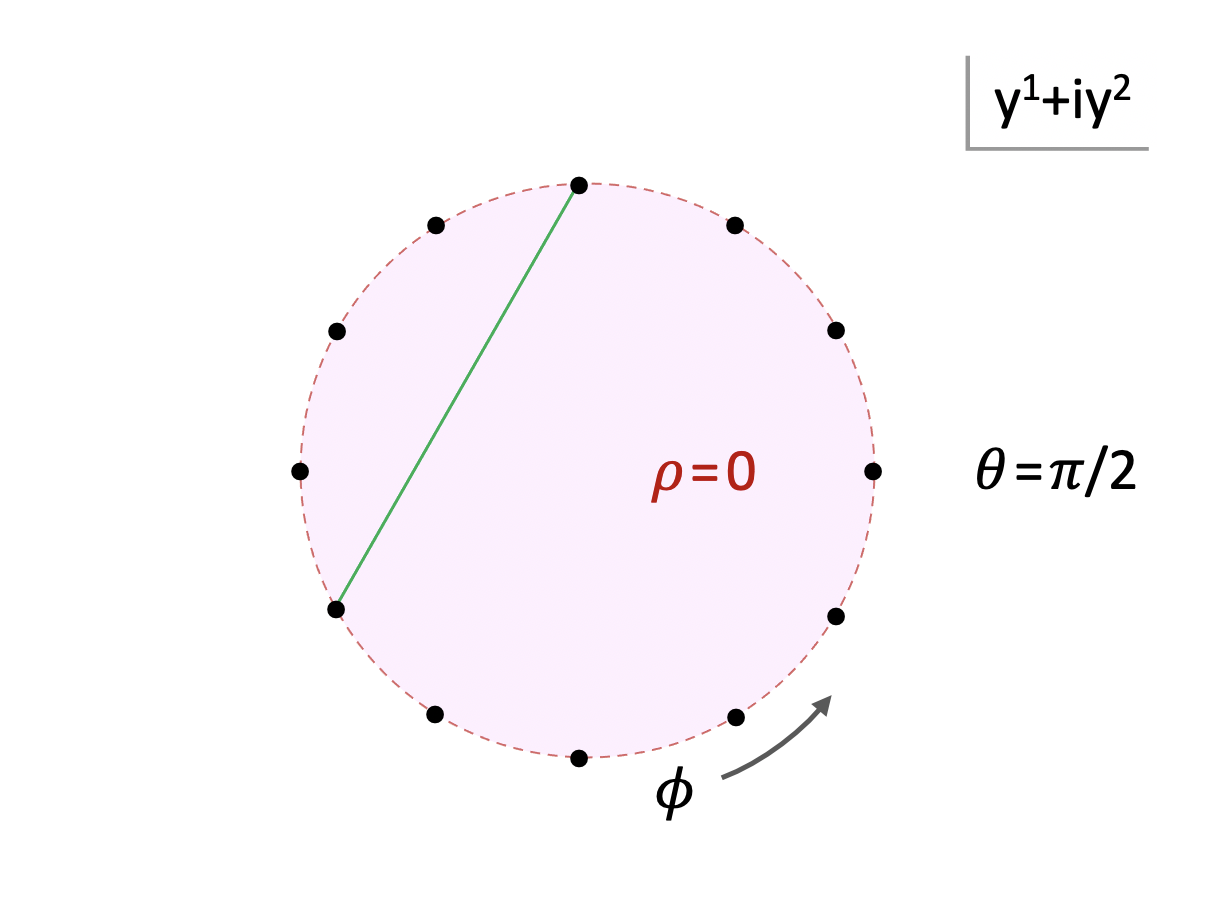}
\caption{\it The gauge projection of a D-brane onto the coset geometry in the $y^1$-$y^2$ plane.  The interior of the ring of fivebranes is the pink disk $\rho=0$, $\theta\in(0,\frac\pi2)$ in the coordinates~\eqref{bipolar}, while the exterior is parametrized by $\theta=\frac\pi2,\rho\in(0,\infty)$.  The W-brane runs along the blue straight line stretched between the locations of NS5-branes uniformly arrayed around the circle that bounds the disk.  For D2-branes in type IIA, the brane must stretch along an additional direction within the fivebrane worldvolume.
}
\label{fig:W-brane}
\end{figure}

Consider first the Coulomb branch fivebranes.  Taking Dirichlet boundary conditions along $x$ and Neumann in $t$, the resulting four-dimensional D-brane state upstairs in $\sltwo\times\sutwo$ loses two dimensions under the gauging (the timelike dimension $\tau$ in $\sltwo$, and the azimuthal direction $\psi$ of the two-sphere in $\sutwo$), and projects down to a brane localized at $\rho=0$, extending along a line in the parafermion disk $\frac\sutwo\uone$, which describes the region in the $y^1$-$y^2$ plane of the geometry interior to the source circle (see figure~\ref{fig:W-brane}).  These straight lines are D-branes stretched along the discrete intervals between fivebrane locations.  If all other spatial directions in the quotient geometry are Dirichlet, one has a D-string stretching between type IIB NS5-branes; if one additional direction is Neumann, one has D2-brane strips in type IIA that are the massive avatars of the little string.  For more details, see~\rcite{Martinec:2019wzw}.    
Along $\bT^4$, the D-brane is free to move, but it is pinned in the transverse space to the location of the fivebranes.  

Moving on to the tilted gauging that yields the supertube of figure~\ref{fig:STsource}, it was shown in~\rcite{Martinec:2019wzw} how to read off the brane geometry in the coset from a choice of conjugacy class branes as well as smearing of those branes along gauge orbits in the underlying group manifolds.  In particular one can see how D2-brane strips in type IIA T-dualize to D3-branes wrapping the KK monopole structure of the NS5-F1 source (see section 7.2 of~\rcite{Martinec:2019wzw}).

It is expected that there will be an equivalent picture of the D-branes in the equivalent FZZ dual superpotential description of the near-source structure (section 6.3 of~\rcite{Martinec:2020gkv} has a qualitative discussion), though the details have not been worked out.

%
%

In the general Lunin-Mathur geometries described in the previous section, D3-branes wrapping the vanishing cycle that appears when the source profile self-intersects, and oscillating along $\bT^4$, become the W-strings of a restored nonabelian little string symmetry at that location.  These D3-branes are T-dual/mirror to the D2-branes stretching between NS5-branes in the NS5-P frame where it is manifest that they are the corresponding W-strings.  The round supertube results of~\rcite{Martinec:2019wzw} were extended to backgrounds where the circular array is deformed to an ellipse in~\rcite{Martinec:2020gkv}; an analysis of the DBI action for the corresponding wrapped D3 branes in the limit that the ellipse degenerates indicates that indeed the tension of the resulting effective string vanishes with the size of the cycle.

Thus, while we can't follow the dynamics into the strong coupling region because we lose control of string perturbation theory, we do indeed see the objects which realize nonabelian little string dynamics as D-branes in weak coupling on separated fivebranes.  The $\half$-BPS configuration space of the NS5-F1 system has limits where subgroups of that nonabelian dynamics are restored, and precisely in those limits we can see in a fully stringy bulk description that the appropriate W-strings are becoming light.  We now wish to repeat this analysis for a class of models that one can call {\it non-critical little string theory}, in which the fivebranes are compactified on a curved manifold such that the decoupling limit loses some of the dimensions that the NS5-branes are wrapping.  The worldsheet description simply doesn't include these dimensions, and so the fundamental string worldsheet theory is in less than ten dimensions, while the little string theory is in less than six dimensions.


\section{Noncritical little string theory}
\label{sec:noncrit}

The above considerations regarding little string theory on $\bS^1_x\times\bT^4$ have a rich extension to little string theory compactified on other geometries, which in many cases admits a solvable worldsheet description for particular $\half$-BPS configurations.
A large class of such solvable worldsheet theories on linear dilaton backgrounds and their $AdS_3$ counterparts was described in~\rcite{Giveon:1999zm,Giveon:1999px,Eguchi:2004ik}.  The linear dilaton theories on the Coulomb branch are noncompact Gepner models 
\be
\label{GKP Coul}
\bT^d \times\bR_t \times \bS^1_x\times
\biggl(\frac{\sltwo_k}{\uone}\times \frac{\sutwo_{\kappa_1}}{\uone}\times\dots\times \frac{\sutwo_{\kappa_r}}{\uone} \biggr)/\bZ_{\kappab}
\ee
with $\kappab=lcm(\kappa_1,...,\kappa_r)$.  The related set of $AdS_3$ backgrounds, obtained by dropping down the linear dilaton throat $\none$ fundamental strings wrapping $\bS^1_x$, is  given by
\be
\label{GKP AdS}
\bT^{d} \times\sltwo_k\times
\biggl({\uone}_x\times \frac{\sutwo_{\kappa_1}}{\uone}\times\dots\times \frac{\sutwo_{\kappa_r}}{\uone} \biggr)/\bZ_{\kappab}  ~.
\ee
In general, these are non-critical string theories~-- one gets roughly two dimensions of target space for each of the group cosets, plus $d+2$ from the rest; the total can be less than ten.  One makes up the difference by adjusting the (continuous) level $k$ of the $\sltwo$ factor.  

We will be interested in the simplest examples $r=1$, and either $d=2$ or $d=0$, which can be thought of as little string theory compactified in the first case on a Riemann surface times $\bT^2$, with the Riemann surface vanishing from the worldsheet description in the decoupling limit; or in the second case, little string theory is compactified on a four-cycle which again disappears from the worldsheet description in the decoupling limit.

As above, a background interpolating between~\eqref{GKP Coul} in the UV and~\eqref{GKP AdS} in the IR is obtained combining all the group factors and gauging null currents~\rcite{Brennan:2020bju}
\be
\label{cosets}
\frac\cG\cH = \frac{\bR_t\times \bS^1_x \times \bT^d \times \sltwo \times \sutwo_\flat}{\uone_L\times \uone_R} ~,
\ee
where the null current lies partly in $\bR_t\times\bS^1_x$ and partly in $\sltwo\times\sutwo$.  In order to maintain spacetime supersymmetry in these non-critical models, the $\sutwo$ WZW model needs to be ``squashed'' by a $J^3_\su\bar J^3_\su$ deformation, which roughly speaking squeezes the size of the $\psi$ circle by a factor $R$ while expanding the $\phi$ circle by a factor $R$ (see~\rcite{Brennan:2020bju} and references therein).  The interpolating geometry is not particularly illuminating, but can be found in section 5 of~\rcite{Brennan:2020bju}.  

The group cosets~\eqref{GKP Coul} are again quantum equivalent to a collection of Landau-Ginsburg models with superpotential $\sfZ_i^{\kappa_i}$ together with an $N=2$ Liouville theory.   In the $r=1$ examples we focus on, the interpretation is pretty much the same as in section~\ref{sec:NS5Coul}~-- a collection of $\nfive=\kappa$ fivebranes separated in their (squashed) transverse space on the Coulomb branch~\rcite{Giveon:1999px}.  This interpretation is natural, given the $\kappa$ units of magnetic $H_3$ flux threading the (squashed) $\bS^3$ in the geometric description obtained by null gauging.    The levels of the various group factors are related by the central charge balance $c_\tot = 15$ of worldsheet string theory
\be
\label{ctot}
\frac{2 c_\tot}3 = 10 = 2+d + \Big(2+\frac 4k\Big) + \sum_{i=1}^r \Big(2-\frac 4{\kappa_i}\Big) ~.
\ee
We consider in detail three cases, all with $r=1$ and $\kappa=\kappa_1=2,3,...$: 
\begin{align}
\label{levelcases}
d=4~&:~~k=\kappa=\nfive ~~;
\nn\\[.2cm]
d=2~&:~~k=\frac{2\kappa}{\kappa+2}~~; 
\\[.2cm]
d=0~&:~~k=\frac{\kappa}{\kappa+1} ~~.  
\nn
\end{align}
We have already discussed the critical dimension case $d=4$ extensively, and so focus more on the non-critical examples $d=2$ and $d=0$.  Both have a rather stringy throat geometry since $k$ is of order one.

The structure of D-branes in these non-critical models is also much the same as in the critical dimension, except that for the D-brane avatars of the nonabelian little string, we have reduced the number of transverse dimensions available for them to wiggle around in.  For the models with $d=2$, a D2-brane wrapped around $\bS^1_x$ and stretching between the fivebranes can wiggle around in $\bT^2$ rather than the $\bT^4$ of the critical dimension.  If this closed path along the D2 worldvolume is not wrapped around something, the brane is unstable and will readily decay.  We can for instance wrap it around one of the cycles of the $\bT^2$ or $\bT^4$.  But even if not, a generically excited collection of D-brane strips will be separated along $\bT^2$ or $\bT^4$ as in figure~\ref{fig:wiggles}; in order to decay the strips must come together so that they may rejoin by condensing the open strings that stretch between them.

\begin{figure}[ht]
\centering
\includegraphics[width=.5\textwidth]{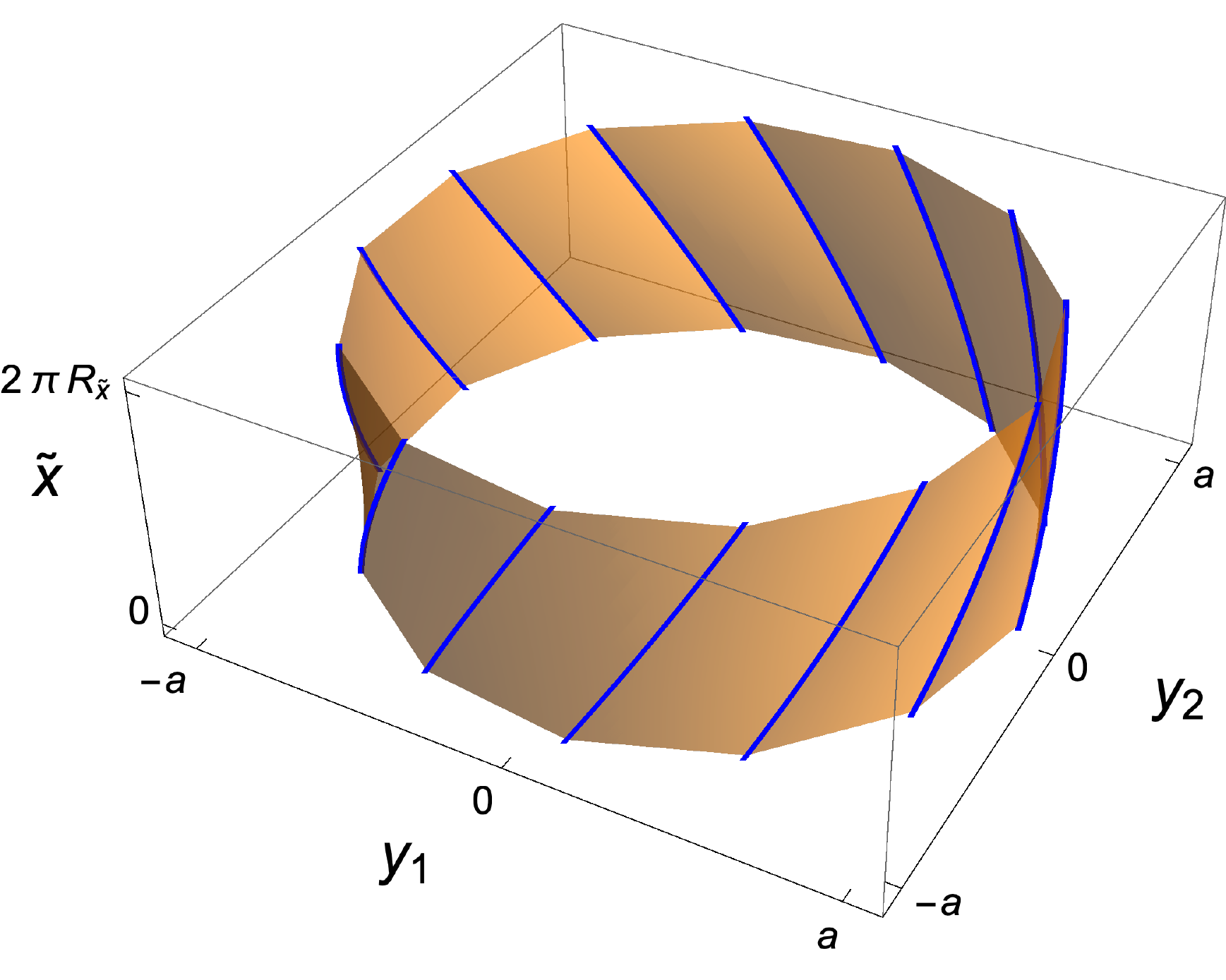}
\caption{\it A W-string (in yellow) that wraps all the way around the NS5 source spiral (in blue).  Since the adjacent segments of the spiral meet at an angle, there is an open string tachyon whose condensation causes the brane to decay.  In the $k<1$ models, there are no additional directions in which the brane segments can separate in order to stabilize the brane against this decay.
}
\label{fig:W-spiral}
\end{figure}

By the time we get to $d=0$ such a D-brane has no light excitations along the fivebrane worldvolume, because there aren't any directions along the fivebranes in the decoupled throat theory besides $\bS^1_x$.  If we are looking to stretch the brane around something to stabilize it, there are no  cycles along the fivebrane other than the spiral in the $x$-$\phi$ cylinder of figure~\ref{fig:STsource}; see figure~\ref{fig:W-spiral}.
This D-brane is always unstable to decay via the open string tachyons that appear where adjacent segments meet at an angle.  The condensate lifts the D-brane away from the fivebranes and allows it to collapse and decay into closed string radiation.  For $d=2$ or $d=4$ the D-branes will also decay eventually, but first they have to find each other in $\bT^2$ or $\bT^4$.
%
%

Thus the D-brane avatars of the little string in the non-critical theories have fewer transverse polarizations available than in the critical dimension~-- only two polarizations for the $d=2$ models, and none for the $d=0$, $k<1$ models of~\rcite{Balthazar:2021xeh}.


\subsection{The correspondence transition at \texorpdfstring{$k=1$}{}}
\label{sec:corrpt}

As argued in~\rcite{Giveon:2005mi} and elaborated in~\rcite{Balthazar:2021xeh}, the $\sltwo$ level $k=1$ marks the dividing line of the string/black hole correspondence principle~\rcite{Horowitz:1996nw}, in that for $k>1$ the high-energy density of states is that of black fivebranes, equation~\eqref{blackNS5}; and in the $AdS_3$ limit $R_x\to\infty$, the high-energy density of states is that of BTZ black holes~\eqref{BTZspec}, where in both these equations we substitute $\nfive\to k$.  In the critical dimension theory on $\bT^4$, one has $k=\kappa=\nfive$; more generally, the density of states is governed by $k$, which is related to the $\sutwo$ levels, and hence any magnetic $H_3$ fluxes present, through the central charge constraint~\eqref{ctot}.  In particular, in the $r=1$ models we consider, it is natural to continue to interpret $\nfive=\kappa$, which differs from $k$ by~\eqref{levelcases}.

One way to see that the black hole spectrum is governed by $k$ rather than $\kappa$ is to examine the linear dilaton regime of the throat.  The black hole entropy in the fivebrane decoupling limit is governed by the Euclidean black NS5 geometry
\be
ds^2 = k\big(\tanh^2\!\rho\, dt^2 + d\rho^2\big) + \kappa\, d\Omega_{\bS^3_\flat}^2 + ds^2_{\bS^1_x} + ds^2_{\bT^d} 
\quad,\qquad
e^{-2\Phi} = \mu\,\cosh^2\!\rho ~.
\ee
where $\mu=\sqrt{k}\,\lstr M$.
The metric in the $\rho$-$t$ plane is that of the $\frac\sltwo\uone$ ``cigar'' coset and this exact solution of the model gives us confidence in the results even when $k\sim 1$.  Feeding this geometry into black hole thermodynamics yields a black hole inverse temperature $\beta_H = \sqrt{k} \,\lstr$, and Hagedorn entropy 
\be
\label{Shag}
S = 2\pi \beta_H M = 2\pi \sqrt{k}\,\lstr \, M ~.
\ee

One can extend these considerations to Euclidean black fivebranes carrying F1 winding charge by modifying the coset to $\frac{\sltwo\times \uone_x}{\uone}$ where again the embedding of the gauge group lies partly in each numerator factor; see~\rcite{Giveon:2005mi}.

FZZ duality again plays a crucial role.  The coset geometry is FZZ dual to $N=2$ Liouville theory, as outlined above.  This means that the near-horizon region at the tip of the Euclidean ``cigar'' geometry comes inextricably linked with a winding string condensate represented by the worldsheet superpotential
\be
\label{windingtach}
\cW = \tilde\mu\, e^{-k(\rho-i\tilde x)}  ~,
\ee
($\tilde x$ is the coordinate T-dual to $x$).  This winding condensate decays exponentially away from the tip of the cigar, and can be interpreted as the Euclidean manifestation of a thermal string atmosphere outside the horizon, in equilibrium with the black hole.

The linearized deformation of the geometry from an asymptotically cylindrical form with linear dilaton, $\delta(ds^2)=e^{-2\rho}dx^2$,  is normalizable in the full geometry.  It corresponds to a worldsheet dimension (1,1) operator in the the $\frac\sltwo\uone$ coset model.%
\footnote{This deformation among other things shifts the value of the dilaton at the tip of the cigar, so one can think of it as the dilaton zero mode.}
But this normalizable deformation of the geometry comes together with the winding condensate~\eqref{windingtach}, and the norm of this string winding component of the deformation 
\be
\int \! d\rho\,dx \sqrt{G} e^{-2\Phi} |\cW|^2 
\ee
is finite for $k>1$ and blows up for $k<1$.  The dimension (1,1) operator includes this stringy contribution and is thus non-normalizable for $k<1$.  The conclusion of~\rcite{Giveon:2005mi} is that because the black hole background itself has such a condensate, it ceases to have a finite action for $k<1$, due to stringy effects, and therefore does not contribute to the thermodynamics.

\begin{figure}[ht]
\centering
\includegraphics[width=.4\textwidth]{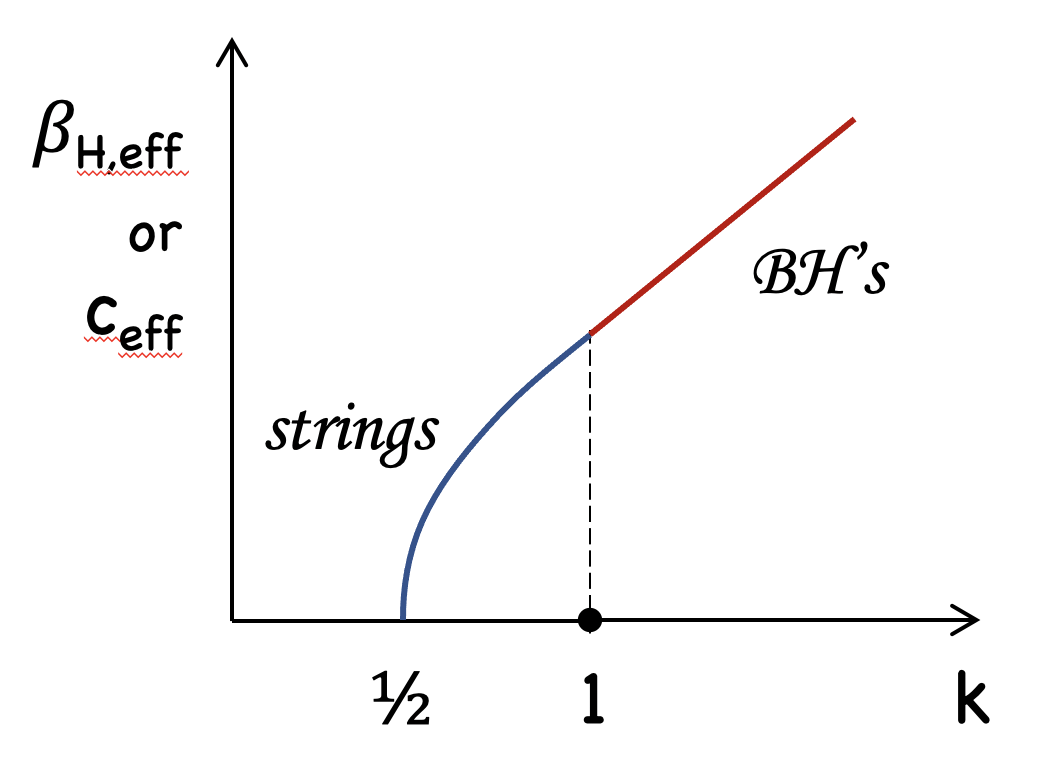}
\caption{\it The effective number of degrees of freedom $c_\eff$ as a function of the $AdS_3$ radius $k$.  Above the correspondence transition at $k=1$ the asymptotic density of states is dominated by the BTZ black hole spectrum; below the transition it is dominated by a gas of fundamental strings.  A similar plot for the effective inverse Hagedorn temperature $\beta_{H,\eff}$ arises in the linear dilaton region, with black fivebranes dominating the entropy for $k>1$ and a perturbative string gas in the linear dilaton throat dominating the entropy for $k<1$.
}
\label{fig:ceff-vs-k}
\end{figure}

So what governs the asymptotic density of states in noncritical little string theory for $k<1$?  For both the regimes with linear dilaton asymptotics and with $AdS_3$ asymptotics, the perturbative string spectrum persists up to arbitrarily high energy and saturates the leading order entropy~\rcite{Giveon:2005mi}.  
One can compute the spectrum of perturbative strings in the linear dilaton region, with the result
\be
S = 2\pi\lstr E\sqrt{2-\frac1k} ~.
\ee
When $k<1$ this entropy is lower than the black fivebrane entropy one would obtain from geometry, equation~\eqref{Shag}, but due to the non-normalizable contribution of the winding condensate the naive geometric entropy formula doesn't apply.  Note however that the two expressions agree at $k=1$. 
At the same point that black holes leave the spectrum, the high energy density of states can be accounted for in perturbative string theory.  
The coefficient $\beta_{H,\eff}$ in the density of states is plotted in figure~\ref{fig:ceff-vs-k}.
%
%

The regime $k<1$ is on the stringy side of the string/black hole correspondence transition, where black holes leave the normalizable spectrum and instead the spectrum consists of a gas of strings and D-branes whose thermodynamics is well-characterized by perturbative string theory.  

What is true of the full little string theory is also true of its CFT limit.  Here one has a spacetime Virasoro algebra, realized on the worldsheet through the string vertex operator amplitude realization of the spacetime CFT correlation functions~\rcite{Kutasov:1999xu,Giveon:2001up}.  The central extension of this algebra arises as the usual leading singularity in the stress tensor OPE
\be
\cT(z_1)\cT(z_2) \sim \frac{3k\cI}{(z_1-z_2)^4} + \frac{2\cT(z_2)}{(z_1-z_2)^2} + \frac{\partial\cT(z_2)}{z_1-z_2} ~,
\ee
but in the worldsheet realization of this algebra, $\cI$ is a nontrivial vertex operator.  This vertex operator has two components, again due to FZZ duality.  One is the zero mode of the dilaton, which again is a normalizable contribution.  The other component is again associated to a winding string condensate; here it represents the $\none$ fundamental strings winding $\bS^1_x$ dissolved in the fivebranes.%
\footnote{More precisely, the parameter $\tilde\mu$ in~\eqref{windingtach} is the chemical potential conjugate to the winding number $\none$~\rcite{Porrati:2015eha}.} 
Just as in the linear dilaton case, the geometrical deformation is always normalizable, but the winding component is non-normalizable for $k<1$.  As a result, there is no normalizable $\sltwo$ invariant state associated to the identity operator, a situation sometimes encountered in non-compact CFT's~\rcite{Seiberg:1992bj}.  But in the torus partition function of the spacetime CFT, the $\sltwo$ invariant state is related by modular invariance to the asymptotic density of states~\rcite{Cardy:1986ie}, which in the holographic context is the BTZ black hole spectrum.  The conclusion of~\rcite{Giveon:2005mi} is that the black hole states also cease to be normalizable.

A more refined computation of the asymptotic density of states in conformal field theories with no $\sltwo$ invariant vacuum yields~\rcite{Kutasov:1990sv}
\be
\label{klt1 ent}
S = 2\pi\sqrt{c_\eff (\varepsilon+p)/2} + 2\pi\sqrt{c_\eff(\varepsilon-p)/2} 
\quad,\qquad
c_\eff = c-24 h_{\rm min}
\ee
where $h_{\rm min}=\bar h_{\rm min}$ is the energy eigenvalue of the lowest energy state in the theory.

There is now also a proposal for the spacetime CFT in the $k<1$ regime~\rcite{Balthazar:2021xeh}, namely a deformation of the symmetric product
\be
(\bR_\rho\times \bS^3_\flat)^\none/S_\none
\ee
where $\bR_\rho$ has a linear dilaton, $\bS^3_\flat$ is the same squashed $\sutwo$ that appeared in the worldsheet construction~\eqref{cosets}, and the deformation is by a $\bZ_2$ twist operator dressed by an exponential of $\rho$ to make it marginal.  This exponential vanishes as $\rho\to+\infty$ where the symmetric product structure reproduces the Fock space of free strings winding $\bS^1_x$, and grows as $\rho\to-\infty$ where it erects a wall to keep the strings out of strong coupling.  It was shown in~\rcite{Balthazar:2021xeh} that the spectrum of this CFT agrees with that of the worldsheet theory on the $d=0$ geometry~\eqref{GKP AdS} for states whose wavefunctions are concentrated in the asymptotic weak-coupling region.

Fundamental strings in purely NS $AdS_3$ and linear dilaton backgrounds have a continuum of winding string states (so-called {\it long strings}) above a gap
\be
h_{\rm min} (w) = \frac{(k-1)^2}{4kw} + \frac k4\Big(w-\frac 1w\Big) 
\quad,\qquad
w= 1,2,...
\ee
The symmetric product reproduces this long string continuum.
This continuum thus has a lowest dimension state where all the strings have $w=1$, leading to a perturbative approximation
\be
\label{klt1 ceff}
c_\eff \approx 6\none\Big(2-\frac 1k\Big) ~.
\ee
The resulting $k<1$ entropy~\eqref{klt1 ent} joins smoothly onto the BTZ entropy with $c=6k\none$ at $k=1$.  

As argued in~\rcite{Balthazar:2021xeh}, this approximation is not expected to have large (\ie\ leading order in $\none$) corrections. 
One can imagine assembling a state by adding strings one at a time from the continuum; when the strings are out at large $\rho$,  the free string approximation accurately calculates their energies in the asymptotically free spacetime CFT elaborated in that work.  The only way this approximation could fail is if there were a bound state spectrum well below the continuum (\ie\ by an amount of order $\none$).  Given that as the energy increases, the wavefunctions spill out to larger and larger $\rho$ where the coupling is weaker and weaker, and thus the perturbative approximation better and better, this possibility seems unlikely.

The three classes of models we chose in order to explore the range of $k$~-- the critical theory on $AdS_3\times \bS^3\times \bT^4$ with $k\ge2$, the $d=2$ models on $AdS_3\times\bS^3_\flat\times\bT^2$ with $1\le k<2$, and the models on $AdS_3\times\bS^3_\flat$ with $k<1$~-- have a similar structure on the worldsheet.  In each case, a worldsheet CFT consisting of a gauged WZW model describes a background that interpolates between a linear dilaton UV and the $AdS_3$ background in the IR.  The $\bT^4$ and $\bT^2$ cases involve $\sltwo$ WZW models with $k\ge 1$, and so have an asymptotic BTZ spectrum; the last class has $k<1$ and so the asymptotic spectrum has no BTZ spectrum.  

The D-brane avatars of the little string have transverse oscillations in the first two cases but not the last, suggesting that when the fivebranes are pushed together and nonabelian little string physics arises in the deep IR, black holes associated to the resulting Hagedorn little string gas will form in the first two cases but not in the last case, because in the last case the little strings have no Hagedorn spectrum.  
This qualitative distinction of D-brane spectra between the three classes of models thus agrees with the general considerations above for the asymptotic spectrum as a function of $k$.  
While the little string naively has a tension lighter than the fundamental string, and hence would be expected to dominate the entropy when it is deconfined, for $k<1$ the little string wrapping $\bS^1_x$ has no transverse oscillations while for $k>1$ it does.  The absence of a black hole spectrum kicks in precisely when the phase space for little string oscillations disappears.


\section{Approximating black holes}
\label{sec:BHapprox}

In this section we discuss features of the $k<1$ models in relation to black holes.  As we have just argued, these models don't have black holes in the spectrum; however, we can approach the black hole threshold by moving through the discretuum of theories accumulating at $k=1$ from below, by taking the level $\kappa$ of the deformed $\sutwo$ in the model to infinity.  We can then ask whether there is a sense in which the properties of the model in the large $\kappa$ limit start to mimic the properties of black holes.  To the extent that the onset of black hole physics arises from a deconfinement transition of nonabelian little strings, and the opening up of an associated large phase space internal to the fivebranes, the answer would appear to be no.  On the other hand, some other features are quite similar, as we shall see.


\subsection{Comments on the asymptotic spectrum}
\label{sec:spec}

The asymptotic spectrum of the $k<1$ models is accounted for by the perturbative string spectrum of the bulk description, extrapolated from low energies and isolated strings to high energies and strings on top of one another.  The energies in question are of order $\none$ or more, in which case the $1/\none$ expansion is suspect.  But the fact that perturbative analysis appears to get the correct asymptotics gives us confidence that it is a reliable guide, at least qualitatively.  Running with this idea, let us piece together a picture of the dynamics in generic high-energy states.

The worldsheet theory has both a discrete spectrum of bound states as well as a continuum of scattering states.  
The bound states are seen as poles in the S-matrix of the scattering states of wound strings that descend into the middle of $AdS_3$ and bounce back out.  At leading order in $1/\none$, the amplitude is diagonal in the winding number $w$ of the string, and there are bound states for every $w$, whose energies are given by the pole structure in the reflection coefficient of the single-string wavefunction.  A multi-string bound state is just a Fock space of such bound strings.
The generic (non-BPS) bound state is not protected by any symmetry,%
\footnote{The BPS bound state spectrum is of course protected by supersymmetry.} 
and can merge with the continuum, \ie\ decay.  The pole locations gain small imaginary parts, and the bound states become resonances which decay by emission of long strings that escape to infinity.

Now, it could be that string perturbation theory is completely breaking down at large energies, and the generic state gains a decay width of order its mass, in which case it no longer makes much sense to talk about a bound state spectrum.  However, we expect the size of the bound state wavefunction to grow with energy and extend further out into the weak-coupling region of the asymptotically free spacetime CFT, in which case string perturbation theory is getting better and better at describing the dominant support of the wavefunction, suggesting that the decay width remains much smaller than the mass (perhaps down by $1/\none$).  The resonant states would then be long-lived, like an ordinary black hole that traps incoming energy and releases it slowly.

Note that as an aside, as the quasi-bound spectrum decays, the late-time state consists of a number of strings heading out toward the boundary $\rho\to\infty$.  But then by time-reversal symmetry the generic state came from the boundary in the far past, by assembling it from an in-state of scattering strings.  To have a steady state in Lorentz signature, it would seem that there would have to be strings all along the $\rho$ direction, both incoming and outgoing, and thus $\none\to\infty$.  One needs to keep feeding strings in from infinity to replenish the ones that are leaking away.%
\footnote{This feature is reminiscent of the $c=1$ matrix model, where the underlying sea of D0-branes (the eigenvalues of the dual matrix model) must be continually fed in, otherwise the sea drains away.}

Once we have dispensed with the radial center-of-mass motion of the long strings by putting them into a quasi-bound state, or by putting the system in a bath of scattering strings, the remaining structure of the Hilbert space seems not all that different from the picture of the $\bT^4$ symmetric product that describes the NS5-F1 system in the critical dimension in a far corner of its moduli space.  That corner has an effective description in terms of which there is only one fivebrane, and the oscillator spectrum of the symmetric product is essentially identical.  Of course, with a single fivebrane, we are again at or near the correspondence transition; nevertheless, the long string spectrum is often taken to be a useful model of black hole dynamics.  To go toward the regimes where an effective supergravity description approximates the bulk dual to the spacetime CFT in the critical dimension, one has to turn on a marginal $\bZ_2$ twist operator in the symmetric orbifold, just as the $k<1$ deformed symmetric product CFT has such a $\bZ_2$ twist operator turned on at finite $\rho$.  Thus we see that there are many points of commonality between the two situations.


\subsection{A picture of dynamics}
\label{sec:evolution}

Another set of interesting questions surrounds the degree to which the evolution of the generic highly excited state in the spectrum approximates that of a black hole, particularly in the $k\to1$ limit, both in its formation and in its decay.  
Consider a situation where most of the CFT is in some low-lying state, for instance one of the BPS ground states discussed in appendix D of~\rcite{Balthazar:2021xeh}, with a probe long string travelling in from large $\rho$ with some radial momentum.
The incoming scattering state in the deformed symmetric product travels toward strong coupling.  Since the wall that protects the strings from strong coupling is a $\bZ_2$ twist operator that joins and splits long strings, the way that a string reflects off the wall is to interact with the condensate of strings already sitting in the IR ensemble.

At low momentum, the worldsheet two-point function predicts that the incident string will reflect off the wall, and return in the same state with a phase shift.  As one dials up the incident $\rho$ momentum, the probe string will climb further up the wall before it reflects, thus it probes further into the strong coupling region~\rcite{Giveon:2015cma}.  The likelihood that it returns unscathed decreases; instead it can transfer some of its energy to long strings in the IR ensemble.  Ideally, the result will match the known winding non-conserving structure of the perturbative string S-matrix (see~\rcite{Dei:2021xgh,Dei:2021yom} for a recent discussion and further references).

At some point, for instance at momenta/energies well above the central charge $c$, the probe string scattering off the IR ensemble will cease to be coherent; instead, the probe string will thermalize into the IR ensemble as it rapidly undergoes many incoherent scatterings deep in the wall.
These scatterings should mediate fast scrambling of the incoming state into the full ensemble of available high-energy states at the incident energy. 
Rapid joining/splitting interactions reshuffle and thermalize segments of long string that make up the state.

Here, the picture is not all that different from that of thermalization in the symmetric product of $\bT^4$ in the critical dimension, except that in the present case the string wavefunction can spread out in $\rho$ to a place where the interactions become weak, whereas in the critical dimension symmetric product the interaction strength is roughly uniform over a compact target space; thus the string wavefunction cannot disperse to a place of weak coupling.  

Let us suppose then that the system is thermalized into a generic excited bound state.
At the surface of such a state, long strings can radiate into the continuum of outgoing modes.  Many aspects of this picture mimic the picture of near-extremal black holes~\rcite{Callan:1996dv,Das:1996wn,Gubser:1997se,Klebanov:1997cx} that foreshadowed the development of gauge/gravity duality.  

The $AdS_3$ models discussed here provide a convenient realization of both the center of $AdS_3$ and a radiation sector, all within a broader CFT or its linear dilaton extension.  The quanta that are capable of leaving the system are in the continuum of wound fundamental strings, with the same sign winding as the background; this will serve as the ``radiation bath'' that the part of the CFT describing the center of $AdS_3$ is coupled to.%
\footnote{Unwound strings lie in bound states rather than in radiation modes.  When one backs away from the decoupling limit, their decaying profiles match onto outgoing spherical waves in the asymptotically flat part of the geometry, which are the usually considered Hawking radiation modes.  But in the strict decoupling limit, the only modes that radiate energy to spatial infinity are the winding strings, assuming that we are working in the locus in moduli space where these strings are not bound but have a continuum of ingoing/outgoing modes.  The models under discussion sit in this locus~\rcite{Seiberg:1999xz,Larsen:1999uk}.} 
The effective field theory Hawking process at the horizon of BTZ black holes in the $k>1$ regime involves the creation of wound-antiwound F1 pairs in the continuum.  The fact that F1 charge is being radiated away can be compensated (if desired) by feeding in unexcited wound strings at an appropriate rate.

The investigation of the thermal radiation from the hot ``lump of string'' that comprises the generic highly excited state of the $k<1$ symmetric product, in particular in the correspondence limit $k\to 1$ where it might start to approximate the Hawking radiation of wound strings, is likely to be a fascinating endeavor.  
%


\subsection{Holographic RG flows}
\label{sec:RGflow}

In~\rcite{Martinec:2020gkv}, the condensation of normalizable $\half$-BPS vertex operators had the effect of deforming the Lunin-Mathur profile function $\sfF(\tilde v)$.  The effect is not to change the theory, but to change the {\it state} in the theory from the $\sltwo$ invariant state to some other $\half$-BPS state.%
\footnote{Note that because the UV theory has $k>1$, the $\sltwo$ invariant state exists in the Hilbert space.}
From the worldsheet point of view, the low-lying modes of deformation correspond to holographic RG flows that deform the background away from the $AdS_3$ vacuum.
One can apply the same logic to non-critical little string models.

In the class of models~\eqref{GKP AdS}, we can consider situations in which we engineer a holographic RG flow between $k>1$ and $k<1$ backgrounds~\rcite{Harvey:2001wm,Balthazar:2021xeh}.  For instance, we can consider the class of models~\eqref{GKP AdS} with $d=0, r=2$.  The associated Landau-Ginsburg potential is 
\be
\cW_0 = \sfZ_1^{\kappa_1}+\sfZ_2^{\kappa_2} - \lambda_0\,e^{-\kappa_1\kappa_2\sfX}
\ee
and the $\sltwo$ level is
\be
\label{ktwo}
k = \frac{2\kappa_1\kappa_2}{2\kappa_1+2\kappa_2+\kappa_1\kappa_2}
\ee
which lies in the range $\frac23\le k < 2$.  Starting from a theory with both $\kappa_1,\kappa_2$ large, one can condense an operator whose worldsheet effect is to flow the superpotential to \eg\ $\kappa_2=2$, in which case the IR theory has $k=\frac{\kappa_1}{\kappa_1+1}<1$.

The associated gauged WZW model of the theory before adding the deformation is~\rcite{Brennan:2020bju} 
\be
\frac{\cG}{\cH} = \frac{\bR_t\times\bS^1_x \times\sltwo\times\sutwo_{\kappa_1}\times\sutwotil_{\kappa_2}}{\uone_L\times\uone_R\times \uone_S}
\ee
where the gauge currents $\uone_{L,R,S}$ are taken to be
\begin{align}
\cJ_L &= J^3_\sl + \sqrt{\frac{k}{2\kappa_1}}\, J^{3,\flat}_\su + \sqrt{\frac{k}{2\kappa_2}}\, J^{3,\flat}_\sutil 
- R_X\big(i\partial t + i\partial X\big)
~~,~~~~
\cJ_S = +J^{3,\flat}_\su - J^{3,\flat}_\sutil
\nn\\[.2cm]
\bar\cJ_R &= \bar J^3_\sl -  \sqrt{\frac{k}{2\kappa_1}}\, \bar J^{3,\flat}_\su -  \sqrt{\frac{k}{2\kappa_2}}\, \bar J^{3,\flat}_\sutil
- R_X\big(i\bar\partial t - i\bar\partial X\big)
~~,~~~~
\bar\cJ_S = -\bar J^{3,\flat}_\su + \bar J^{3,\flat}_\sutil
\end{align}
The first two currents $\cJ_L,\bar \cJ_R$ are null, while $\cJ_S,\bar\cJ_S$ is a standard (spacelike) axial gauging.  The squashing factors of $\sutwo,\sutwotil$ are respectively
\be
R_i = \sqrt{\frac{\kappa_i}{2k}}
~~,~~~~
i = 1,2~,
\ee
in order that the background is $N=(2,2)$ supersymmetric.  The normalizable $\half$-BPS vertex operators include 
\begin{align}
\label{Wtil}
{\widetilde\cW}_{j_1',j_2'}^{(-1)}=
e^{-\varphi-\bar{\varphi}} \,
 \,
e^{i\frac{(Y+\bar{Y})}{\sqrt{2k}}(j-k/2)} \,
\Lambda^{\left(0,0\right)}_{j_1';j_1',j_1'} \,
\widetilde\Lambda^{\left(0,0\right)}_{j_2';j_2',j_2'} \,
\Phi^{(-1)}_{j;j,j}\,  ~.
\end{align}
The notation is as follows:
\begin{itemize}
\item
$Y$ here parametrizes the physical combination of the scalars that bosonize $J^3_\su,J^3_\sutil$ left over after eliminating $U(1)_S$;
\item
$\Lambda^{(\alpha,\bar\alpha)}_{j_1';m_1',\mbar_1'}$ are operators in the $\frac\sutwo\uone$ superparafermion theory in the sector $(\alpha,\bar\alpha)$ of R-symmetry spectral flow (similarly for $\widetilde\Lambda^{(\alpha,\bar\alpha)}_{j_2';m_2',\mbar_2'}$ and $\frac\sutwotil\uone$);
\item
$\Phi^{(w)}_{j;m,\mbar}$ are primaries of winding $w$ and spin $j$ in the bosonic $\sltwo$ WZW model.
\end{itemize}
See~\rcite{Brennan:2020bju,Balthazar:2021xeh} for further details; our conventions follow~\rcite{Balthazar:2021xeh}.
The $\sutwo$ spins take the values $j_i'=0,\half,...,\half\kappa_i-1$, and
\be
j= k\Big(\half - \frac{j_1'}{\kappa_1} - \frac{j_2'}{\kappa_2} \Big)
~~,~~~~
p_Y = \frac{j-k/2}{\sqrt{2k}}
~~,~~~~
p_X = - \bar p_X = 0 ~.
\ee
For $j_1',j_2'$ such that
\be
\label{normable}
\frac{\kappa_1\kappa_2-2j_1'\kappa_2-2j_2'\kappa_1}{\kappa_1\kappa_2-2\kappa_1-2\kappa_2} > \half ~.
\ee 
these vertex operators lie in the range of the unitary discrete series of normalizable $\sltwo$ operators
\be
\half<j<\frac{k+1}2  ~,
\ee
which correspond to normalizable string states in the $AdS_3$ cap.  Let $\cN_1$ bet the set of such normalizable operators with $w=-1$.  There are additional normalizable $\half$-BPS (NS,NS) sector vertex operators in higher winding sectors $w<-1$~\rcite{Argurio:2000tb,Martinec:2020gkv,Balthazar:2019rnh}, but they are explicitly $t$-dependent while the operators above are $t$-independent.
One can also consider the non-normalizable operators either inside or outside the range~\eqref{normable}.  
For operators inside the normalizable range, this corresponds to turning on a chemical potential for the corresponding states; operators outside the range deform the theory rather than the state in a given theory.

When added to the action these normalizable vertex operators describe the deformation of the background to other states in the $\half$-BPS state space of the theory at large $\none$, just as in the critical dimension the condensates of the $\half$-BPS operators shift the Lunin-Mathur profile function $\sfF(\tilde v)$, and lead to a deformed supergravity solution as sketched above (see~\rcite{Martinec:2020gkv} for details).  

On the worldsheet, condensing these lowest operators corresponds to implementing an RG flow in the Landau-Ginsburg model, dressed by $\sltwo\times\uone$ in the $AdS_3$ regime, and worldsheet dilaton gravity in the linear dilaton regime.  In the closely related Coulomb branch gauging where we drop the $\partial t,\partial X$ terms in the null currents, these deformations are Coulomb branch moduli that move the underlying fivebranes around in a correlated way, by deforming the Landau-Ginsburg potential via
\begin{align}
\cW &= \cW_0+\delta\cW
\nn\\[.2cm]
\delta\cW &=
\sum_{j_1',j_2'\in\cN_1} \lambda_{j_1',j_2'}^{(1)} \sfZ_1^{2j_1'} \sfZ_2^{2j_2'} e^{-(\kappa_1\kappa_2 - 2\kappa_2 j_1' - 2\kappa_2 j_2')\sfX} ~.
\end{align}
The corresponding deformations~\eqref{Wtil} of the supertube geometry deform the background by pushing around the wiggly fivebrane source configuration, by changing the state of the condensate of fundamental strings dissolved in the fivebranes.

In particular we can engineer a flow to an effective $k<1$ regime using a $\sfZ_1$ independent deformation $\delta\cW$ that pushes the fivebranes apart in the $\sfZ_2$ plane, leaving a $k<1$ effective theory at low energies.  While the bulk effective geometry is currently unknown, it would seem to have the structure of an asymptotically linear dilaton background with the ultraviolet value of $k$~\eqref{ktwo}, which is sourced in the IR by a state whose local geometry is governed by a dynamics with an effective $k\approx \frac{\kappa_1}{\kappa_1+1}$ below some scale $\rho^*$ in the radial direction (this scale can be larger or smaller than the F1 charge radius, depending on our choice for $R_x$).

We are thus in a situation where the high-energy density of states is governed by BTZ black holes whose horizon radius $\rho_h > \rho^*$, but the low-energy density of states consists of a gas of Hagedorn strings, until the size of the ball of string reaches $\rho\sim\rho^*$.  The transition between these two behaviors is the correspondence transition of~\rcite{Horowitz:1996nw}.

We can thus approach the correspondence transition from above, by having the horizon evolve through a background geometry that reflects the holographic RG flow through the correspondence transition.  For black holes above the transition, the horizon radius is larger than the holographic correspondence scale, and the BTZ solution is valid.  The little string has Hagedorn entropy and dominates the density of states..  However, FZZ duality dictates that the Euclidean black hole background automatically comes together with a winding string condensate along the lines of~\eqref{windingtach}; in the $AdS_3$ regime, this condensate is reflected in the FZZ dual of the identity operator.  In Lorentz signature, this condensate is believed to analytically continue into a fundamental string gas in equilibrium with the black hole.  As the mass decreases, nonabelian little string excitations gradually freeze out, and disappear at the correspondence point.  Below the correspondence point, black holes leave the spectrum and one is left with just the string gas.    

It is the inability to localize the fundamental string that is responsible for its soft UV behavior in string perturbation theory; the absence of a singularity and the absence of a horizon come together on the stringy side of the correspondence transition.  It is tempting to speculate that the nonabelian little string plays the same role on the black hole side of the transition, both resolving the singularity and having a wavefunction sufficiently delocalized that there is in the end no horizon as well as no singularity in the fundamental description (even if the effective description exterior to the black hole has such a horizon, and analytically continues to an interior geometry with such a singularity).


\section{Wormholes, islands, and the correspondence transition}
\label{sec:EPRnotER}

The $k<1$ models provide insights into the nature of the string/black hole transition, and the nature of holography in stringy regimes more generally.  In this section we draw two conclusions: First, that the absence of a BTZ solution means that entanglement in stringy regimes of holography does not always create a wormhole, no matter how strong the entanglement is; and second, that the when the correspondence transition occurs in the course of black hole evaporation, it can distinguish between the island and fuzzball scenarios for the structure of the black hole interior by allowing us to peek inside.


\subsection{Connective dissonance}
\label{sec:ivermectin} 

A major difference between NS5-F1 models with $k<1$ and the NS5-F1 system in the critical dimension is that one has solid arguments that for $k<1$ the BTZ black hole solution does not govern the asymptotic spectrum.  It does not describe any sort of ensemble average over high-energy states; nor does it describe typical members of that ensemble (using the eigenstate thermalization hypothesis).  For $k<1$ the BTZ black hole is not normalizable, nor is its extension to the linear dilaton regime.  Yet the picture of such states in the spacetime CFT in terms of a symmetric product is much the same.
It is then interesting to revisit some accepted lore about gauge/gravity duality in light of the results discussed above.

Note that the bulk picture of the $k<1$ models~-- to which we have much more access in the spacetime CFT than is usually the case in gauge/gravity duality because the radial coordinate in the bulk is part of the dual field theory~-- is much more reminiscent of the fuzzball proposal.  There are no horizons present, and the states are simply complicated bound states of strings and branes.

The leading order entropy in the $AdS_3$ limit still has Cardy-like growth~\eqref{klt1 ent}.
At large $\kappa$, \ie\ $k=\frac{\kappa}{\kappa+1}\to1^-$, the coefficient $c_\eff$ of equation~\eqref{klt1 ceff} approaches the central charge $c=6\none k$, and thus the entropy approaches the BTZ entropy.

\begin{figure}[ht]
\centering
\vspace{-0.2cm}
  \begin{subfigure}[b]{0.45\textwidth}
    \includegraphics[width=\textwidth]{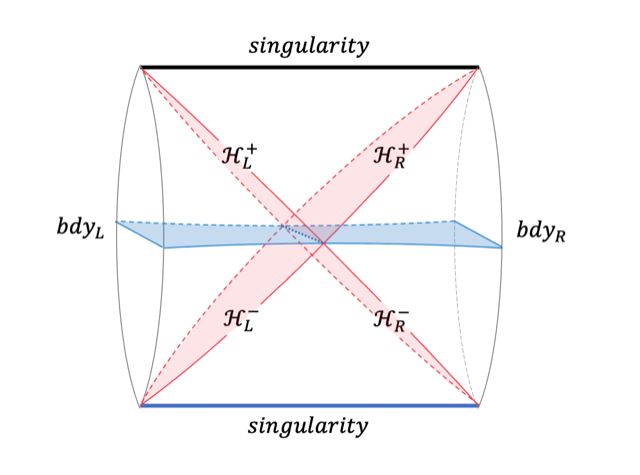}
    \caption{ }
    \label{fig:BTZpenrose}
  \end{subfigure}
    \vspace{0.1cm}
  \begin{subfigure}[b]{0.45\textwidth}
    \hskip -0.3cm
    \includegraphics[width=1.1\textwidth]{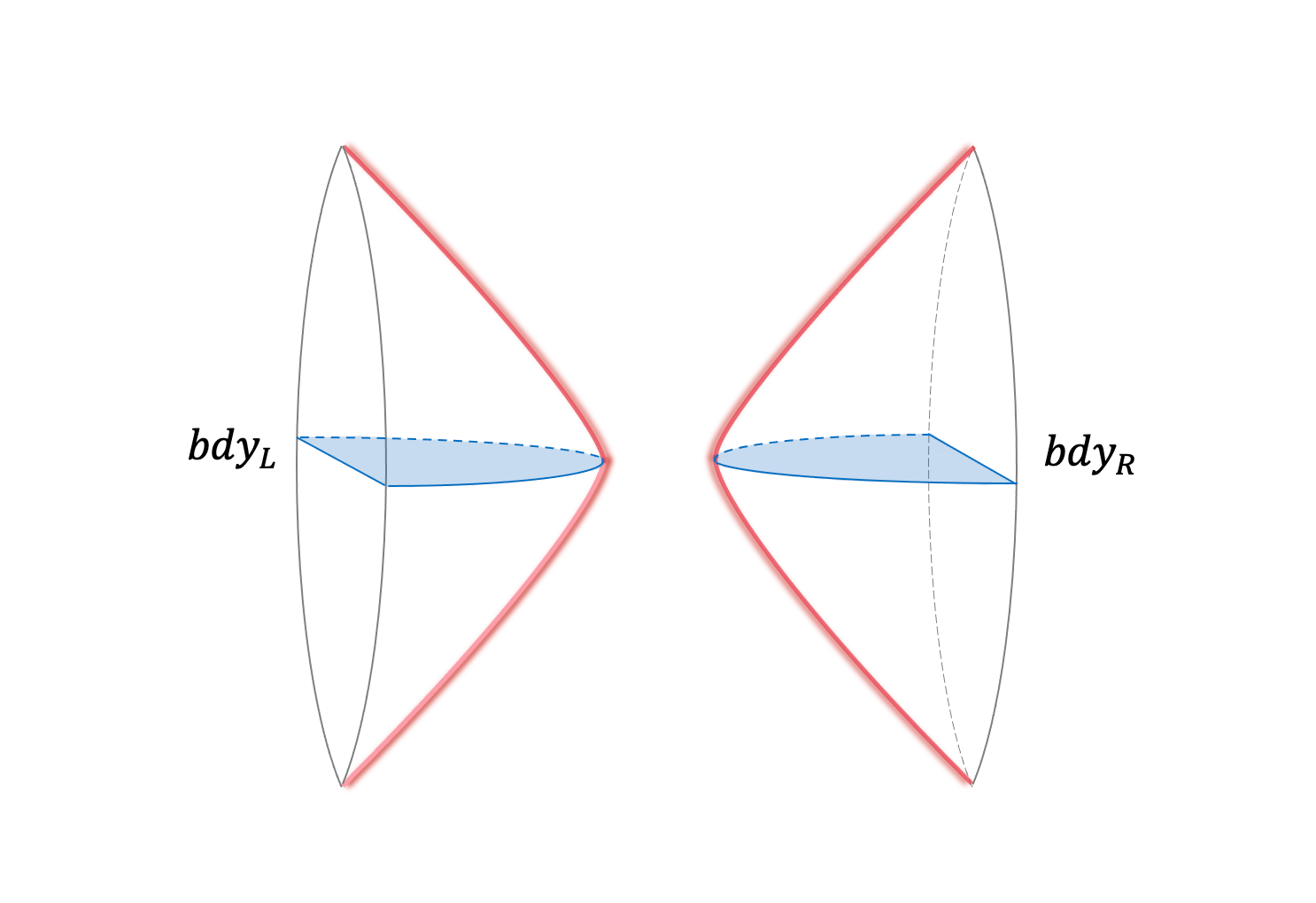}
    \caption{ }
    \label{fig:gasPenrose}
  \end{subfigure}
\caption{\it 
(a) Penrose diagram of the eternal BTZ black hole.  The diagram is periodically identified front to back, so that the spatial sections (\eg\ the blue surface) have cylindrical topology.  The horizons $\cH^\pm_{L,R}$ causally separating the two asymptotic regions are shown in red.
(b) Penrose diagram for the thermofield double of a thermal AdS gas.
}
\label{fig:sources}
\end{figure}

A standard construction in thermal field theory is the thermofield double state in the tensor product Hilbert space of two copies of a QFT
\be
\label{double}
\big|{\it TFD}\big\rangle =  
\sum_{\ket{\alpha}\in\cH_{\rm CFT}} e^{-\beta E_\alpha/2} \,\ket{\alpha}_1\ket{\alpha}_2 ~,
\ee
for instance the norm squared of this state is the thermal partition function.
In holographic theories where the high-energy spectrum is dominated by black holes, this state is thought to be dual to the eternal black hole geometry~\rcite{Maldacena:2001kr};
the Penrose diagram of the eternal BTZ black hole is depicted in figure~\ref{fig:BTZpenrose}.
%

For $k<1$, the thermofield double in the deformed symmetric product CFT instead describes an entangled Hagedorn gas of wound strings, depicted in figure~\ref{fig:gasPenrose}.  While the low-energy spectrum for $k>1$ is also such a gas of quanta in $AdS_3$, for $k>1$ there is a Hawking-Page transition to the black hole phase~\rcite{Witten:1998zw}, which as we have argued above is associated to a deconfinement transition of nonabelian little strings.  But according to the analysis of~\rcite{Giveon:2005mi} reviewed above, for $k<1$ the associated BTZ solution does not correspond to a normalizable state.  Instead one has the correlated AdS gas of figure~\ref{fig:gasPenrose}, which governs the thermodynamics up to arbitrarily high temperatures.

A theme in recent research concerns a proposed relation between quantum entanglement and geometry~\rcite{Maldacena:2001kr,VanRaamsdonk:2010pw}~-- that entanglement {\it creates} geometry, such as the regions beyond the horizons in figure~\ref{fig:BTZpenrose}.  But here we have a situation where one can consider a highly entangled state such as~\eqref{double}, but the usually associated semclassical geometry does not seem to exist~-- there is never a Hawking-Page transition because there is no phase dominated by nonabelian little strings.  
Could there still be some stringy saddle point that replaces figure~\ref{fig:BTZpenrose}, and which characterizes aspects of the thermofield double~\eqref{double}, and yet still has the properties of a wormhole geometry (\eg\ with a slightly smaller area at the neck of the surface of time symmetry to reflect the somewhat reduced density of states)?  If so, then the horizon of that saddle is not reflective of the existence of a horizon in any member of the ensemble, because black holes are not normalizable states.
Instead, entanglement does not seem to create (albeit stringy) wormholes under these circumstances (as has for instance been argued in~\rcite{Mathur:2014dia}).  On the stringy side of the correspondence transition there are no causal horizons, and so a stringy version of figure~\ref{fig:BTZpenrose} does not appear.
This observation naturally raises the question of whether such geometrical structures (horizons, wormholes) develop in the exact theory as one passes through the correspondence transition, or whether they are also absent on the other side of the transition in the exact theory, as has also been argued for instance in~\rcite{Mathur:2014dia}.  This could be the case if for instance little strings at $k>1$ play the same role that fundamental strings do for $k<1$.

The $k<1$ models seem to provide a useful laboratory in which to investigate these issues.
In particular, in the context of the ``entanglement creates geometry'' idea it is sometimes said that any minute amount of entanglement generates a tiny (Planck-size) wormhole connecting the entangled degrees of freedom.  Instead, here we have a situation where there can be an arbitrarily large amount of entanglement, and the purported wormhole geometry does not apply.


\subsection{Juxtaposing the correspondence transition and the Page time}
\label{sec:ivermectin} 

Consider now the dynamics of black holes that during the course of evaporation encounter the correspondence transition.  We will be interested in the interplay of the timing of this transition relative to the Page time.

A black hole with horizon radius $\rho_h$ above the correspondence transition scale $\rho^*$, decays until it reaches the correspondence transition at $\rho_h\sim\rho^*$.  During this transition, the horizon of the effective field theory description melts away.  At this point, the black hole becomes a highly excited string gas, and whatever secrets it was hiding are revealed.  

There are many ways we can engineer such a transition.  
One of the cleanest would be to manually dial the gauge theory coupling constant of the $N=4$ $SU(N_c)$ supersymmetric gauge theory dual to $AdS_5\times\bS^5$, coupled to an external bath so that a black hole in $AdS_5$ can evaporate.
In the bulk, this procedure would send a coherent pulse of the dilaton in from the AdS boundary, whose profile can be made sufficiently adiabatic that one is not sending an enormous shock wave toward the black hole, but instead simply making the string scale $\lstr$ larger relative to $R_{AdS}$, holding $R_{AdS}/\lpl$ fixed.  As $R_{AdS}/\lstr$ decreases to less than one, a black hole of arbitrary mass becomes a weakly coupled SYM plasma.  Similar considerations apply to the D1-D5 system dual to $AdS_3$, where again $g_6^{\,2}\equiv\gstrsq/{{V_{\bT^4}}}$ is a modulus.

In the F1-NS5 models we have been discussing, a holographic RG flow from $k>1$ in the UV to $k<1$ in the IR generates a throat that has string scale curvature at a radial scale $\rho^*$ and below; decaying black holes whose horizon radius decreases to this value will undergo a correspondence transition.
It is unlikely, though, that one can tune $\rho^*$ to be arbitrarily large relative to the $AdS$ or the string scale; instead, one expects that horizon radius of order $\rho^*$ corresponds to a mass of order the central charge.  In this respect, the manual dialing of the string coupling holds the advantage of being able to arrange the correspondence transition to occur when the black hole has any mass, not just a mass of order the central charge or less.

The considerations of the previous subsection suggest that an evaporating thermofield double passing through the correspondence transition makes a transition of semiclassical saddles between figure~\ref{fig:BTZpenrose} and~\ref{fig:gasPenrose}, in which the wormhole pinches off while the system still has a large entanglement entropy between the two sides.  This needs to happen in such a way that the BTZ wormhole doesn't become traversible, since there is no quantum channel operating between the two sides, just entanglement.

One may now ask about the interplay between the correspondence transition and various proposed resolutions of the information problem.  We will restrict our attention to two such proposed resolutions: The island scenario and the fuzzball scenario.  We will focus mostly on the former, then comment briefly on the latter.

\begin{figure}[ht]
\centering
\includegraphics[width=1.0\textwidth]{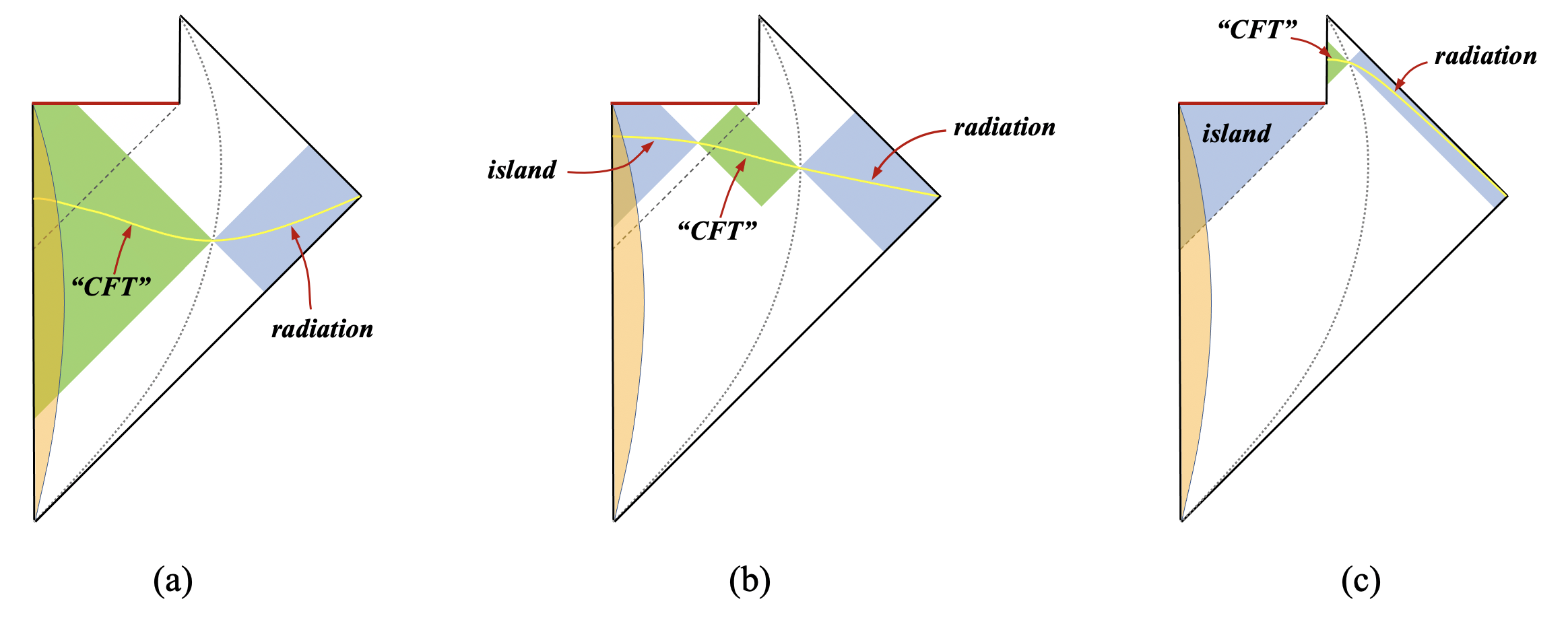}
\caption{\it The standard Penrose diagram for black hole formation and evaporation, showing different Cauchy slices (yellow) in the island scenario.  The matter forming the black hole occupies the region in orange.
(a) Initially, the black hole is described by low energy CFT modes usually associated with the center of $AdS_3$, whose entanglement wedge is shown in green (we can for instance take the outer boundary of this region to be the F1 charge radius, where the geometry rolls over to the linear dilaton fivebrane throat).
(b) During evaporation, an island forms in the black hole interior; degrees of freedom there are encoded in the entanglement wedge of the radiation, shown in blue.  
(c) The aftermath of the evaporation process, with the island left behind.}
\label{fig:islands}
\end{figure}

In the island scenario (see for instance~\rcite{Almheiri:2020cfm} for a review and further references), the conventional Hawking process occurs at the smooth horizon of a black hole in an effective theory of gravity that describes the black hole evolution.  In this effective gravity theory, there exist smooth Cauchy slices that contain the partners of emitted Hawking quanta, as well as the incoming matter that formed the black hole, and in the semi-classical approximation these slices can hold an arbitrary amount of independent low-energy degrees of freedom.  Such a slicing is depicted in figure~\ref{fig:islands}.
%
%

\begin{figure}[ht]
\centering
\includegraphics[width=0.5\textwidth]{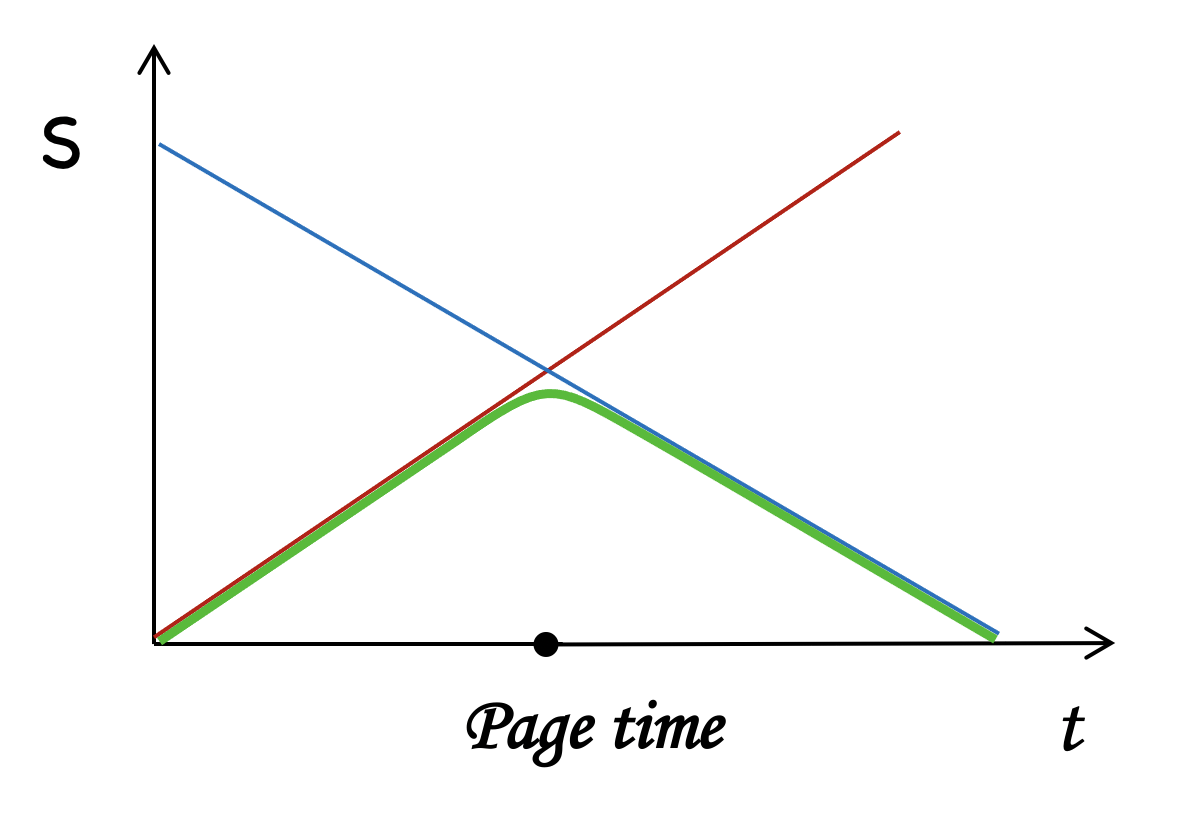}
\caption{\it The entropy of Hawking radiation (red) monotonically grows; the Bekenstein-Hawking entropy given by the black hole horizon area (blue) monotonically decreases.  By associating the island portion of the black hole interior to the radiation, the entropy associated to the black hole follows the Page curve (green).}
\label{fig:page}
\end{figure}

The initial matter that formed the black hole as well as the entangled partners of early Hawking quanta inhabit an {\it island} on the Cauchy slice, whose boundary is determined by an extremization procedure (and hence is called a {\it quantum extremal surface}).  To resolve the information paradox, it is postulated that this island becomes ``part of the radiation", as depicted in figure~\ref{fig:islands}b, starting at the Page time (roughly halfway through the evaporation process); in this way, degrees of freedom ostensibly inside the black hole are transferred to the radiation state space through some unspecified mechanism long after the Hawking quanta they are entangled with have left the vicinity of the black hole.  After the black hole has evaporated, what was the black hole interior is encoded entirely in the radiation Hilbert space, as shown in figure~\ref{fig:islands}c.
By declaring the island to be part of the radiation, the entropy associated to the black hole follows the Page curve in figure~\ref{fig:page} (depicted in green).
%
%

\begin{figure}[ht]
\centering
\includegraphics[width=0.7\textwidth]{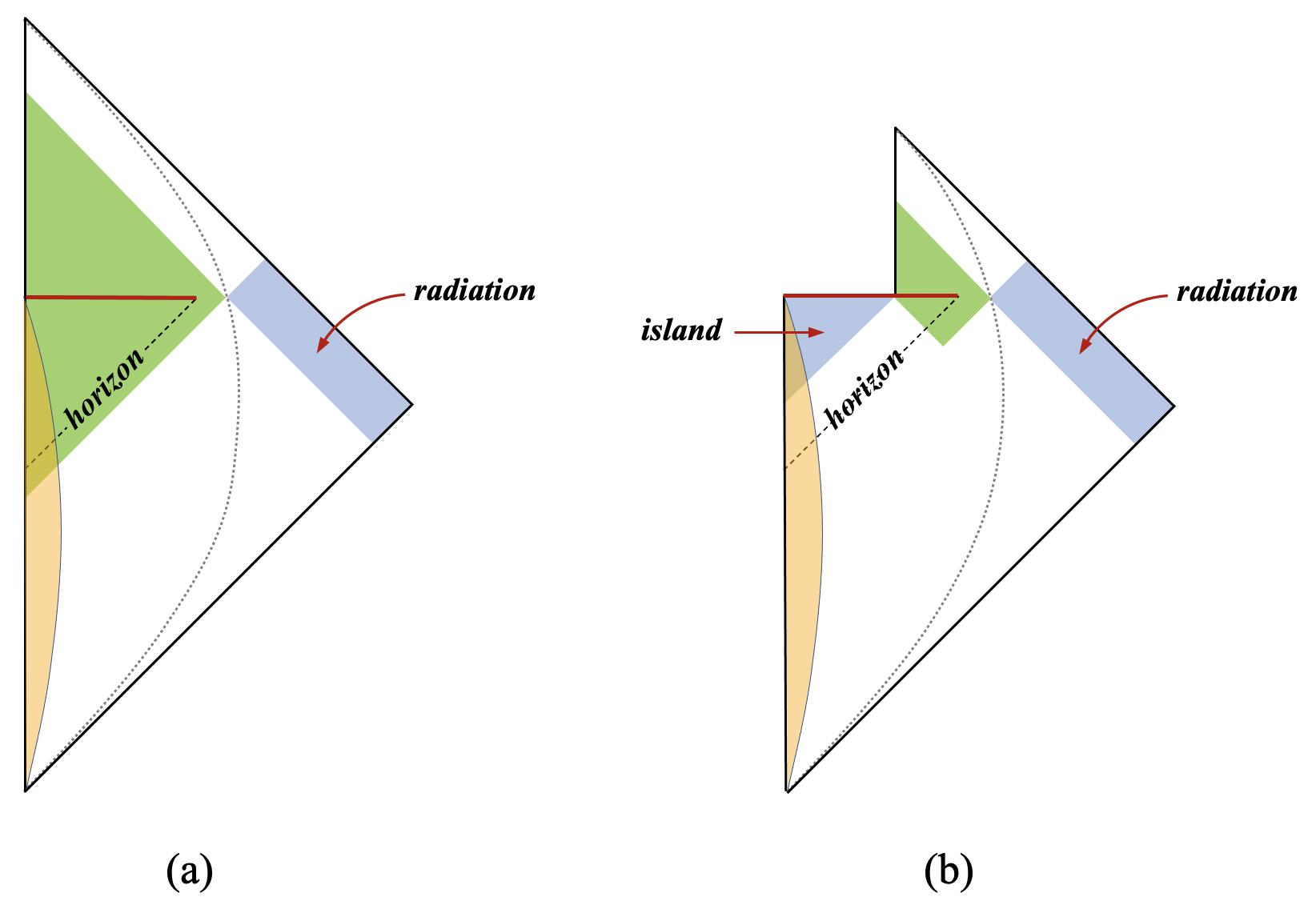}
\caption{\it The Cauchy slice at the time of the correspondence transition.  
(a) When the transition occurs before the Page time, the entire black hole interior is in the entanglement wedge of the degrees of freedom describing the center of $AdS_3$.
(b) When the transition occurs after the Page time, there is an island in the entanglement wedge of the radiation; the radiation state space contains the degrees of freedom describing the entangled partners of the early radiation generated by the Hawking process. }
\label{fig:island-transition}
\end{figure}

We now wish to inject a new ingredient into the discussion, by arranging for the correspondence transition to occur at some point during the evaporation process (though we note that there is an inherent imprecision about this of order the string scale $\lstr$ due to string fuzziness).  We can arrange for the evaporation process to reach the Page time either before or after the correspondence transition.  These two possibilities lead to two different pictures of the transition:
\begin{enumerate}[i)]
\item
If the black hole encounters the correspondence point first, before the Page time, the black hole interior is supposed to still lie in the entanglement wedge of the CFT degrees of freedom describing the center of $AdS$.  Then the entire black hole interior is in the entanglement wedge of the CFT at the time of the transition, and so should be revealed in the string gas post-transition.  This outcome is depicted in figure~\ref{fig:island-transition}a. 
\item
If, however, the evolution passes through the Page time first, then a substantial part of the black hole interior lies in the entanglement wedge of the radiation state space.  The black hole interior depicted in figure~\ref{fig:island-transition}b has an island.  When the correspondence transition then occurs, spacetime bifurcates at the quantum extremal surface which separates the entanglement wedge of the center of $AdS_3$ from the island, with the island pinching off and separating from the rest of spacetime and only the green region becoming a stringy gas visible to the exterior. 
\end{enumerate}
Either way, after the correspondence transition, the remaining lump of string radiates via conventional string joining/splitting interactions.  One would then have a situation where the late radiation is or is not correlated to the early radiation, depending on whether the correspondence transition occurs before or after the Page time.

It seems that there is a discontinuous jump between these two pictures when the Page time and the correspondence transition coincide.  Is the black hole interior part of $AdS$, or part of the radiation?  This discontinuity arises from the insistence that low-energy effective field theory describe the black hole interior, so that the conventional Hawking process takes place at the horizon.  We must then have somewhere to store the entangled partners of the Hawking quanta.  This storage device is the part of the smooth Cauchy slice in the black hole interior.  In island holography, one ascribes the degrees of freedom encoding these quanta to either the CFT or radiation state space according to whose entanglement wedge they lie in.  This wedge jumps abruptly at the Page time as two saddles in the generalized entropy formula exchange dominance.  The correspondence transition, though, must reveal a large portion of the interior, and thus whether the partner quanta are still accessible or not to the CFT.  In figure~\ref{fig:island-transition}a the answer is yes; in figure~\ref{fig:island-transition}b the answer is no.

The two saddles describe distinct realizations of the black hole interior prior to the correspondence transition.
One saddle has a trivial quantum extremal surface, and encodes the entire black hole interior within the CFT state space.  In particular, the interior partners of the previously emitted Hawking quanta, and the infalling matter that formed the black hole are so encoded.  The second saddle declares (when we are in the black hole phase) that these interior partners are now part of the radiation Hilbert space, along with all the rest of the interior of the quantum extremal surface.  This surface lies near the would-be horizon, and encodes the entanglement of the CFT with the island.

If the interior is revealed via the correspondence transition before the Page time, the entire black hole interior lies in the entanglement wedge of the CFT, and therefore the information within the black hole is entirely encoded in the CFT; that interior is entangled with the radiation because it contains the Hawking partners.  This is the picture in figure~\ref{fig:island-transition}a.   If the interior is revealed after the Page time, the CFT remains highly entangled with the radiation state, but now this entanglement is geometrical entanglement across the neck (located at the quantum extremal surface) of some wormhole geometry.  But we have argued in the previous subsection that such geometries don't exist on the stringy side of the correspondence transition. 
Then the neck of the spatial geometry at the quantum extremal surface should pinch off, and the geometrical entanglement across that surface then becomes ordinary quantum entanglement between the remaining string gas and the now disconnected island.  This is the picture in figure~\ref{fig:island-transition}b.
%
%

But while in this post-Page time correspondence transition the CFT is entangled with the island, and the Hawking partners on the island are entangled with the Hawking radiation, these are two different sorts of entanglement between two entirely different sets of degrees of freedom~-- one is geometrical entanglement across the quantum extremal surface, which after the transition becomes ordinary entanglement between the string gas and the island; while the other is entanglement of low-energy effective QFT degrees of freedom between the Hawking quanta and their island partners.  Due to the monogamy of entanglement, there cannot be any entanglement between the CFT degrees of freedom in the remaining string gas and the early Hawking quanta.  This lack of entanglement between the early and late radiation (the latter coming from the string gas) should be contrasted with the situation when the transition occurs before the Page time, where entanglement wedge reconstruction implies that the remaining string gas in the CFT will be entangled with the early radiation, and therefore the early and late radiation will be entangled.

We can also imagine cycling the system through the correspondence transition in both directions.  For instance, turn the black hole into a string gas somewhat before the Page time, so that the interior lies in the entanglement wedge of the CFT.  Perhaps let the system radiate a bit or otherwise interact with its environment, and then dial up the coupling so that it becomes a black hole again, but with less than half the mass of the original one.  Let that Hawking evaporate until somewhat before the Page time of this smaller black hole, then dial through the correspondence transition; and so on.  In the end the system has evaporated completely, for most of the time as a black hole, and yet in the end nothing has disappeared into an island after which it is inaccessible to the outside observer.

Thus there seems to be some tension between the island formalism as usually stated, and the correspondence transition.  The apportionment of spacetime has a discontinuous jump at the Page time, where a portion of the black hole interior jumps from being part of the CFT to being part of the radiation.  Correspondingly the degrees of freedom located on that part of the slice jump from being part of the CFT to being part of the radiation.  This jump is ordinarily invisible to the external observer since that part of the Cauchy slice is hidden behind an event horizon, but the correspondence transition opens the black box and allows us to peer inside and see what's there, and what's not.

The fuzzball paradigm provides a rather different scenario for black hole dynamics.
In the present context, the black hole in this scenario is a fuzzball composed of nonabelian little strings bound to NS5-branes, and one does not have an effective geometrical description of the interior, at least insofar as the Hawking process is concerned.%
\footnote{While much of the recent literature on fuzzballs is devoted to exploring various horizonless solutions of the classical effective supergravity theory (see~\rcite{Shigemori:2020yuo} for a review), the considerations in sections~\ref{sec:critdim}-\ref{sec:Dp} above paint a picture where such geometries can evolve along near-BPS trajectories to singular configurations, at which point the stringy degrees of freedom responsible for black hole entropy are liberated.  The generic fuzzball is then expected to be non-geometrical and dominated by these entropic degrees of freedom.}
In the black hole regime, rather than the radiation arising through a process of string pair creation, instead fractionated little strings assemble themselves and ``unfractionate'' into a fundamental string leaving the black hole bound state (these fundamental strings form the thermal atmosphere associated to the Euclidean winding condensate).  There is no horizon in the fundamental theory, no pair creation at that horizon and thus no partner string behind it; black holes radiate conventionally.%
\footnote{If there is any sense in which the conventional picture of Hawking pair creation holds, it might be along the lines of condensed matter systems in which the absence of a quasi-particle excitation can be construed as a quasi-hole, and quasi-particle/quasi-hole pair creation in which the quasi-particle is extracted from the system leaves behind an entangled quasi-hole, but fundamentally there is no quasi-hole but rather just a modified state of the underlying constituents.}
The radiated strings are built directly out of the entropic degrees of freedom of the black hole microstate, whose wavefunction is supposed to be coherent across the horizon scale.  The information paradox is resolved in this scenario by a direct entanglement of the radiation with the black hole microstate left behind, rather than having it entangled with some partner string through a vacuum pair creation process, uncorrelated with the state of the black hole.   
Every emitted string delivers information to the external observer, regardless of when it was emitted.  The late-emitted quanta are always entangled with the early-emitted quanta, regardless of the ordering of the Page time and the correspondence transition.


\vskip 1cm


\section*{Acknowledgements}

We thank 
B. Balthazar, 
A. Giveon,
D. Kutasov,
S. Mathur,
and
M. Rangamani
for useful discussions.
This work is supported in part by DOE grant DE-SC0009924.


\appendix


\vskip 3cm

\bibliographystyle{JHEP}      

\bibliography{fivebranes}


\end{document}
